\documentclass[a4paper,11pt]{article}
\usepackage{slashed}
\usepackage{cancel}
\usepackage{color}
\usepackage{amssymb}
\usepackage{amsmath}
\usepackage{amsfonts}
\usepackage{graphicx}
\usepackage{xcolor}
\newtheorem{lemma}{Lemma}

\newtheorem{theorem}{Theorem}

\def\cP{{\bf \check P}}
\def\LL{{\cal L}}

\def\pr{{\rm pr\,}}
\def\mod{{\rm mod\ }}
\def\wtg{{\tilde g }}
\def\wtx{{\tilde x }}
\def\haf{{1 \over 2}}

\def\d{\mathrm{d}}

\def\u{{\chi}}

\def\z{{\zeta}}

\def\bC{{\Bbb C}}
\def\bR{{\Bbb R}}

\def\bZ{{\Bbb Z}}
\def\Im{\mathop{\rm Im}\nolimits}
\def\Re{\mathop{\rm Re}\nolimits}
\def\Arg{\mathop{\rm Arg}\nolimits}

\def\tg{\mathop{\rm tg}\nolimits}
\def\th{\mathop{\rm th}\nolimits}
\def\ch{\mathop{\rm ch}\nolimits}
\def\sh{\mathop{\rm sh}\nolimits}

\def\wt{\widetilde}
\def\ovl{\overline}

\def \vhi{\varphi}
\def \veps{\varepsilon}

\def\z{{\zeta}}
\def\Im{\mathop{\rm Im}\nolimits}
\def\Re{\mathop{\rm Re}\nolimits}

\def\tg{\mathop{\rm tg}\nolimits}
\def\th{\mathop{\rm th}\nolimits}
\def\ch{\mathop{\rm ch}\nolimits}
\def\sh{\mathop{\rm sh}\nolimits}

\def\CC{{\cal C}}

\def\LL{{\cal L}}
\def\MM{{\cal M}}
\def\NN{{\cal N}}

\def\RR{{\cal R}}

\def\TT{{\cal T}}
\def\UU{{\cal U}}
\def\VV{{\cal V}}

\def\wt{\widetilde}
\def\ovl{\overline}

\def \vhi{\varphi}
\def \veps{\varepsilon}

\def\half{{\scriptstyle{1 \over 2}}}
\def\interior#1{\setbox1=\hbox{$#1$}\rlap{$#1$}\kern0.4\wd1\raise1.1\ht1%
\hbox{$\scriptstyle \circ$}}
\def\bydef{\mathrel{\buildrel \hbox{\scriptsize \rm def} \over =}}
\def\boxit#1#2{\setbox1=\hbox{\kern#1{#2}\kern#1}%
\dimen1=\ht1 \advance \dimen1 by #1 \dimen2=\dp1 \advance \dimen2 by #1
\setbox1=\hbox{\vrule height\dimen1 depth\dimen2\box1\vrule}%
\setbox1=\vbox{\hrule\box1\hrule}%
\advance \dimen1 by .4pt \ht1=\dimen1 \advance \dimen2 by .4pt \dp1=\dimen2
\box1\relax}
\def\endprf{\raise .5ex\hbox{\boxit{2pt}{\ }}}
\def\rT{{\rm T}}

\def\u{\chi}

\def\ifundefined#1{\expandafter\ifx\csname#1\endcsname\relax}
\def\ttheta{\zeta}
\def\beq{\begin{equation}}
\def\endq{\end{equation}}
\def\eeq{\end{equation}}
\def\beqa{\begin{eqnarray}}
\def\bea{\begin{eqnarray}}
\def\endqa{\end{eqnarray}}
\def\eea{\end{eqnarray}}
\def\x{{\theta}}

\def\m{{l}}

\def\P{{\bf P}}
\def\Q{{\bf Q}}

\def\z{{\zeta}}

\def\tg{\mathop{\rm tg}\nolimits}
\def\th{\mathop{\rm th}\nolimits}
\def\ch{\mathop{\rm ch}\nolimits}
\def\sh{\mathop{\rm sh}\nolimits}

\def\wt{\widetilde}
\def\ovl{\overline}

\def \vhi{\varphi}
\def \veps{\varepsilon}

\def\slr{SL(2, \bR)}
\def\slc{SL(2, \bC)}

\def\ifundefined#1{\expandafter\ifx\csname#1\endcsname\relax}

\def\beq{\begin{equation}}
\def\endq{\end{equation}}
\def\beqa{\begin{eqnarray}}
\def\endqa{\end{eqnarray}}
\def\twofcn{{W}}

\renewcommand{\cosh}{\ch}\renewcommand{\sinh}{\sh}
\renewcommand{\tanh}{\th}\renewcommand{\tan}{\tg}
\def\dateline{\today}
\date{\dateline}

\usepackage{amssymb}

\title{\bf QFT and topology in two dimensions: ${\bf \slr}$-symmetry and the de Sitter universe}
\author{Henri Epstein$^{1}$ and Ugo Moschella$^{2}$\\
$^{1}$Institut des Hautes Etudes Scientifiques (IHES), \\ 35, Route de Chartres,  91440 Bures-sur-Yvette, France\\
$^{2}$Universit\`a dell'Insubria, DiSat, Via Valleggio 11, 21100 Como\\
and INFN sezione di Milano, Via Celoria 16, Italia}
\date{\dateline}

\begin{document}
\maketitle

\begin{abstract}
We study bosonic Quantum Field Theory on the double covering $\widetilde{dS}_{2}$ of the 2-dimensional de Sitter universe, identified to a coset space of the group $\slr$. The latter acts effectively on $\widetilde{dS}_{2}$ and can be interpreted as it relativity group. The manifold is locally identical to the standard the Sitter spacetime ${dS}_2$; it is globally hyperbolic, geodesically complete and an inertial observer sees exactly the same bifurcate Killing horizons as in the standard one-sheeted case. The different global Lorentzian structure causes however drastic differences between the two models. We classify all the $\slr$-inveriant two-point functions and show that: 1) there is no Hawking-Gibbons temperature; 2) there is no covariant field theory solving the Klein-Gordon equation  with mass less than $1/2R$ , i.e. the complementary fields go away.
\end{abstract}


\section{Introduction}

Quantum field theory on curved spacetime provides to date the most reliable access to the study of quantum effects when gravity is present. 
The procedure amounts to finding a metric solving  Einstein's equations and then studying quantum field theory in that background, possibly considering also the back-reaction of the fields on the metric in a semiclassical approach to quantum gravity. The Hawking effect \cite{hawking1,hawking2} and the spectrum of primordial perturbations \cite{CM}  are found and characterized in this way.

There is however a caveat : by solving the Einstein equations one gets the local metric structure of the spacetime but the global topological properties remain inaccessible.
This problem is old and well-known (but nowadays a little under-appreciated) in cosmology: the question of the overall shape and size of the universe and of its global features dates back to the ancient philosophers and becomes a theme with Aristotle, Nicholas of Cusa and Giordano Bruno. This sort of  study is today  a research field that goes under the name of cosmic topology (see e.g \cite{Luminet,LR}). 

The global topological properties may also play an important role in the interplay between quantization and causality : they do and cannot be bypassed.
Two very well known examples of this status of affaires are the G\"odel universe and the anti-de Sitter spacetime which have closed timelike curves and the light-cone ordering is only local. In the anti de Sitter case a cheap way to "treat" problem (which by the way is not so much challenging) consists in moving to the universal covering of the manifold :  this  allows for scalar theories of any mass to be considered.
Another well known quantum topological effect  is the Aharonov - Bohm phenomenon: the presence of a solenoid makes the space non-simply connected.

An example of a similar nature that we consider in this paper is based on the two-dimensional de Sitter spacetime which is topologically non trivial in the spacelike directions; here we study scalar quantum fields on its double covering.

Two-dimensional models of quantum field theory have played and still play an important role as theoretical laboratories where to  explore quantum phenomena that have then been recognized to exist also in four dimensional realistic models, the best known beeing the Schwinger\footnote{For example the Schwinger model, which corresponds to two-dimensional quantum electrodynamics, has allowed for the pre-discovery of some of the most important phenomena expected from quantum chromodynamics such as asymptotic freedom and confinement.
} \cite{sch} and the Thirring \cite{thir,kla} models. 
By considering the same models formulated on the two-dimensional de Sitter universe  ${dS_2}$ one may ask what are the features that survive on the curved manifold. This study  may also throw new light on some of the difficulties encountered in perturbation theory. 

To start this program we have recently reconsidered the free de Sitter Dirac field in two dimensions, founding interesting new features \cite{Epstein1} which are related to  the  two inequivalent spin structures of the two-dimensional de Sitter manifold. Correspondingly, there are two distinct Dirac fields which may be either periodic (Ramond) or anti-periodic (Neveu-Schwarz) w.r.t. spatial translations (rotations in the ambient space) of an angle $2\pi$. A requirement of de Sitter covariance (in a certain generalized sense) may be implemented at the quantum level only in the anti-periodic case \cite{Epstein1,difrancesco}. 
As a consequence,  the Thirring-de Sitter model admits de Sitter covariant solutions \cite{Epstein2} only in the antiperiodic case.   
The double covering of the de Sitter manifold $\widetilde{dS_2}$ naturally enters in the arena of soluble models of QFT through that door.

The manifold $\widetilde{dS_2}$ is in itself a complete globally hyperbolic manifold. It carries a natural {\em effective and transitive} action of $\slr$, the double covering of  $SO_0(1,2)$, the pseudo-orthogonal de Sitter relativity group that acts on ${dS_2}$. The Lorentzian geometry of $\widetilde{dS_2}$ is locally indistinguishable from that of ${dS_2}$ but the global properties are quite different. This fact has striking consequences at the quantum level; some of them have already been announced in \cite{bmds2}. In this paper we give a full characterization of the massive Klein-Gordon fields of $\widetilde{dS_2}$ {\em by constructing the most general local and $\slr$-invariant two-point functions.} The most surprising  physical consequnces are the following :  there exists no analogue of the  so-called Bunch-Davies vacuum \cite{Bunch,Chernikov,hawking,bm,bgm,bem} on the double covering of the de Sitter spacetime and therefore the Hawking-Gibbons  \cite{hawking,bm,bgm,bem}  temperature disappears; furthermore there are no nontrivial truly $\slr$  - invariant (as opposed to $SO_0(1,2)$- invariant) states whose mass is less than $m_{cr} = 1/2$. 

It is known that thermal effects in QFT are related to the existence of a bifurcate Killing horizon. 
Examples of such
spacetimes include Minkowski spacetime, the  extended Schwarzschild spacetime  and
de Sitter spacetime. Kay and Wald \cite{kayw} have proven a uniqueness theorem 
for such thermal states but they also provide counterexamples  \cite{kayw}  where such a  geometrical structure does not imply the existence of the corresponding thermal state
as in the Schwarzschild-de Sitter and in the Kerr cases.

Our example is in a sense more  peculiar: the double covering of the two-dimensional manifold is indistinguishable for a geodesic observer  
from the uncovered manifold. There is no classical experiment that can be done do to determine whether he or she  lives in the de Sitter universe or its double covering.
Yet at the quantum level things are different and the global geometric structure of the double covering makes $\slr$  invariance, locality and  the relevant analyticity properties of the correlation functions
incompatible and this forbids the existence of the thermal radiation from the horizons. One final lesson that may be drawn from our simple example is that we may perhaps probe the global shape of the universe by quantum  experiments made in our laboratories.

\section{The de Sitter universe  as a coset space.}
\label{ds2coset}

Let us consider the two-dimensional de Sitter group $G=SO_0(1,2)$
which is the component  connected to the identity of 
the  pseudo-orthogonal Lorentz group acting on the three-dimensional Minkowski spacetime $M_3$ with metric   $(+,-,-)$.
The Iwasawa decomposition $KNA$ of a generic element $g$ of $G$ is written as follows:
\begin{eqnarray}  g &= &k(\ttheta) n(\lambda)  a(\u)  = 
\exp (\zeta {\bf e_k}) \ \exp (\lambda  {\bf e_n})\  \exp  (\chi  {\bf e_a})  = \cr  && \cr
&=& \begin{array}{l} \label{iwa}
 \left(
 \begin{array}{rrr}
 1 & 0 & 0 \\
 0 & \cos \ttheta & \sin \ttheta \\
 0 & -\sin \ttheta & \cos \ttheta \\
\end{array}
\right)  
\left(
\begin{array}{ccc}
1+ \frac{\lambda ^2}2 & -\frac{\lambda ^2}{2} & \lambda  \\
 \frac{\lambda ^2}{2} & 1-\frac{\lambda ^2}{2} & \lambda  \\
 \lambda  & -\lambda  & 1 \\
\end{array}
\right)
 \left(
\begin{array}{rrr}
 \cosh \chi & \sinh\chi & 0 \\
 \sinh  \chi& \cosh \chi & 0 \\
 0 & 0 & 1 \\
\end{array}
\right). 
\end{array}
\end{eqnarray}
The above decomposition gives natural coordinates $(\lambda, \ttheta) $ to points  
$x= x(\lambda, \ttheta)=k(\ttheta)n(\lambda)$ 
of the coset space $G/A$,   which is seen to be topologically a cylinder.
 Here $\ttheta$ is a real number $\mod 2\pi$. 
 
 Once chosen the coset representatives $x(\lambda, \zeta) $, the left action of the group $G$ on $G/A$ is explicitly written as follows:  
 \begin{equation}
 x'= x(\lambda',\ttheta') = g \  x(\lambda,\ttheta)\  a(g,x(\lambda, \zeta)) \label{trasf1a}
 \end{equation}
The case of a rotation $k(\beta) \in K$
is of course the easiest to account for and amounts simply to a shift of the angle $\ttheta$:
$
\lambda' (\beta) =  \lambda ,\ 
 \ttheta' (\beta)= \ttheta+\beta,
$
where both $\ttheta$ and $\ttheta'$ are real numbers $\mod 2\pi$. The two other subgroups  give rise to slightly more involved transformation 
rules.\footnote{ \label{fooy}A boost $a(\kappa)\in A$ gives\begin{eqnarray}\left. \begin{array}{l}  \lambda'(\kappa) = \lambda  \cosh \kappa+\sinh \kappa (\lambda  \cos \ttheta+\sin \ttheta), \  \sin\ttheta' (\kappa)=\displaystyle{\frac {\sin \ttheta }{\cos \ttheta  \sinh \kappa +\cosh \kappa }},\ \cos\ttheta' (\kappa)=\displaystyle{\frac{\cos \ttheta  \cosh \kappa +\sinh \kappa }{\cos \ttheta  \sinh \kappa +\cosh \kappa }}.\end{array}\right.\label{trasf2a} \end{eqnarray}
 %
 An element $n(\mu)\in N$  gives \begin{eqnarray}\left. \begin{array}{l}   \lambda' (\mu)= {\lambda} \left(1 +\frac 12 {\mu ^2}\right)- \mu  \left( \lambda +\frac{\mu}2 \right) \sin \ttheta+{\mu } \left(1- \frac12 \lambda  \mu \right) \cos \ttheta, \\\sin\ttheta'(\mu) =\displaystyle{ \frac{2 \sin \ttheta -2\mu (1-\cos \ttheta )}{\mu ^2(1- \cos \ttheta )+2(1- \mu  \sin \ttheta )}},\ \ \cos\ttheta'(\mu)= \displaystyle{\frac{\mu ^2 (1-\cos \ttheta )-2 \mu  \sin \ttheta +2 \cos \ttheta }{\mu ^2(1- \cos \ttheta )+2(1- \mu  \sin \ttheta )}}.\end{array}\right.\label{trasf3a}\end{eqnarray}}

By introducing  the variables
\begin{equation}
u= \tan \left( \frac{\ttheta }{2}+\arctan \lambda \right),  \ \ \ \ v = \cot \frac{\ttheta }{2},\label{variables}
\end{equation}
the action of $g(\alpha,\mu,\kappa)= k(\alpha)n(\mu)a(\kappa)  \in G$ becomes interestingly simple: 
\begin{equation}
u\to u'= \frac{(e^\kappa u +\mu)\cos \frac{\alpha}{2}+\sin \frac{\alpha}{2}}{\cos \frac{\alpha}{2}-(e^\kappa u +\mu) \sin\frac{\alpha}{2}} ,\ \ 
 v\to  v'= \frac{(e^\kappa v - \mu) \cos \frac{\alpha}{2}-\sin \frac{\alpha}{2}}{\cos \frac{\alpha}{2}+(e^\kappa v -\mu)  \sin
 \frac{\alpha}{2}}. 
  \end{equation}
The group action on the coset space transforms the variables $u$ and $v$ homographically. 

The Maurer-Cartan 1-form $g^{-1}{\rm d} g $
provides a left invariant metric on $G/H$ as follows:
\begin{eqnarray}
x^{-1}{\rm d} x  &=& \text{d$\zeta $}\, {\bf e_k} +  \left(\frac{1}{2}  \lambda ^2 \d\zeta +\text{d$\lambda $} \right)\,{\bf  e_n }  +  \lambda   \text{d$\zeta $} \, {\bf e_a}  
= \cr &=& \omega 
 + (  \lambda   \text{d$\zeta $}) \ {\bf e_a}   \label{pop}
\end{eqnarray}
 Eqs. (\ref{trasf1a}) and (\ref{pop}) then imply that
 \begin{eqnarray}
{x'}^{-1}{\rm d} x'= a^{-1} ( x^{-1}{\rm d} x) \, a +   a^{-1} \d a  = 
 a^{-1} \,\omega\, a+    (  \lambda   \text{d$\zeta $}) \ {\bf  e_a } +  a^{-1} \d  a \label{PPP}
\end{eqnarray}
Therefore 
 \begin{equation}
\d s^2 =   \frac 12  \makebox{Tr} (\omega^2) =  - (\lambda ^2+1) {\d \zeta^2 }-2 {\d\lambda
   } {\d\zeta }.
   \label{dscov}
\end{equation}
is invariant under the action (\ref{trasf1a}). 
Note that the Iwasawa coordinate system $(\lambda,\ttheta)$ is not orthogonal and  that  $\d s^2$ {\em  is not } the restriction to the submanifold $\chi = 0$  of the Maurer-Cartan metric  
 \begin{equation} \frac 12 {\rm Tr} (g^{-1} \d g\  g^{-1} \d g) = -\text{d$\zeta $}^2-2 \text{d$\zeta $} \text{d$\lambda $}+2 \text{d$\zeta $} \text{d$\chi $} \lambda
   +\text{d$\chi $}^2.  
   \end{equation}

Changing to the variables $u$ and $v$ the metric takes the form
\begin{equation}
\d s^2= \frac{4\, \text{d}u \, \text{d}v}{(u+v)^2};
\end{equation}
$u$ and $v$ are light-cone coordinates and the metric is conformal to the Minkowski metric.
 
A simple geometrical interpretation of the above construction  may be unveiled by introducing the standard representation of $dS_2$ as a one-sheeted hyperboloid
\begin{equation}
dS_{2}=\left\{ x \in {M}_{3}:\ {x^{0}}^{2}-{x^1}^{2}-{x^2}^{2}=- 1 \right\} .\label{ds}
\end{equation}
The coset space $G/A$ can indeed  be identified with  $dS_{2}$ as follows: given a point $x\in G/A$ let us associate to it a vector in ${M}_{3} $ whose components 
are the entries $x_{02},x_{12}$ and $x_{22}$ of the third column of the matrix $x(\lambda,\zeta)$ which  is obviously invariant by  the right action of the subgroup $A$;  one also has that $x_{02}^2- x_{12}^2-x_{22}^2=-1$.   
The so-defined  map is a bijection between $G/A$ and $dS_2$  and  gives  natural global coordinates $(\lambda, \ttheta) $ to points\footnote{We adopt the same letter $x$ to denote points of the coset space $G/A$ and of the de Sitter hyperboloid $dS_2$, as they are identified.}  of the de Sitter hyperboloid:
\begin{equation}
x(\lambda,\ttheta)= \left\{
\begin{array}{l}
x^0= \lambda , \\
x^1= \lambda  \cos \ttheta+\sin \ttheta ,\\
x^2= \cos \ttheta-\lambda  \sin \ttheta .\\
\end{array} 
\right.\label{pariwa}
\end{equation}
The left action of $SO_0(1,2)$ on the coset space $G/A$ by construction coincides with the linear action of  $SO_0(1,2)$ in $M_3$ restricted to the manifold $dS_2$:
$$x(\lambda',\ttheta') = g x(\lambda,\ttheta)$$ 
and the metric coincides with the restriction of the ambient spacetime metric to de de Sitter manifold
\begin{equation}
{\d}s^2 = \left.\left({\d x^0}^2-{\d x^1}^2-{\d x^2}^2\right)\right|_{dS_2} = -2 \d\lambda \d\ttheta  -\left(\lambda ^2+1\right)\d\ttheta^2. \label{dsmet}
\end{equation} 
 The base point (origin) $x(0,0)= (0,0,1)$ is invariant under  the action of the subgroup $A$.
Of course (\ref{pariwa})
supposes that we have chosen a certain Lorentz frame in $M_3$, and
this frame will remain fixed in the sequel.

The light-cone variables $u$ and $v$ have  a simple geometric interpretation; they correspond to the two ratios that may be formed by  factorizing the equation defining the de Sitter hyperboloid: 
\begin{eqnarray}
u = \frac {1-x^2}{x^{1}-x^0} 
= \tan \left( \frac{\ttheta }{2}+\arctan \lambda \right) \ \ \ \ 
v = \frac {1+x^2}{x^{1}-x^0}
= \cot  \frac{\ttheta }{2}
\end{eqnarray}
In term of the light-cone variables we get the following parametrization of the de Sitter manifold
\begin{equation}
x(u,v)= \left\{
\begin{array}{l}
x^0= \frac{u v-1}{u+v}, \\
x^1= \frac{u v+1}{u+v},\\
x^2= \frac{v- u}{u+v}.\\
\end{array} 
\right.\label{pariwa2}
\end{equation}

The complexification of the de Sitter manifold can  be equivalently identified either with the coset space $G^c/A^c$ of the corresponding complexified groups  or with the complex de Sitter hyperboloid 
\begin{equation}
dS_2^c = \{ z \in {\bf C}^3 : {z^0}^2-{z^1}^2-{z^2}^2= -1\}.
\end{equation}
As a complex 2-sphere, $dS_2^c$ is simply connected.
Particularly important subsets of $dS_2^c$ are the forward and backward tuboids, defined as follows \cite{bm,bgm,bem} :
\begin{eqnarray}
{\cal T}^+ = \{ z \in dS_2^c : {(\Im z^0)}^2-{(\Im z^1)}^2-{(\Im z^2)}^2> 0, \ \ \Im z^0>0 \},\\
{\cal T}^- = \{ z \in dS_2^c : {(\Im z^0)}^2-{(\Im z^1)}^2-{(\Im z^2)}^2> 0, \ \ \Im z^0<0 \}.
\end{eqnarray}
These two domains can also be shown to be simply connected.

In the following we will make also use of the standard global orthogonal coordinate system:
\begin{equation}
x(t,\theta)= \left\{
\begin{array}{l}
x^0= \sinh t , \\
x^1= \cosh t   \sin \theta ,\\
x^2= \cosh t \cos  \theta.\\
\end{array} 
\right.\label{parglo}
\end{equation}
Here $\theta$ is a real number $\mod 2\pi$. The relation between the two above coordinate system is quite simple:
\begin{eqnarray}
\lambda = \sinh t, \ \ \tan \theta = \tan(\ttheta +\arctan \lambda). \label{idi}
\end{eqnarray}

\section{The double covering of the 2-dim de Sitter manifold as a coset space.}
\label{coset}
The easiest and most obvious way to describe the double covering  $ \widetilde{dS_2}$  of the two-dimensional de Sitter universe  ${dS_2}$ consists in unfolding the periodic coordinate $\theta$ in (\ref{parglo}). More precisely, we may write the covering projection
$\pr: \widetilde{dS_2} \rightarrow dS_2$ as follows:
\begin{equation}
\pr (\tilde x (t,\theta)) = x(t,\theta\ \mod 2\pi), \label{proju}
\end{equation}
where we use the coordinates $(t,\theta)$  to parameterize  also $\widetilde{dS_2}$; at the lhs $\theta$ is a real number $\mod 4\pi$.

The double covering $ \widetilde{dS_2}$ arises also as a coset space of the double covering $\widetilde{G}=\slr$ of $SO_0(1,2)$.
An element 
\begin{equation}
\wtg= 
\left(
\begin{array}{cc}
 a &  b \\
 c & d \\
\end{array}
\right)
\label{sp12}
\end{equation}
of $\wt G$ is parametrised by four real numbers $a,b,c,d$ subject to the condition  $\det \tilde g = a d-bc=1.$   For $\tilde g \in \wt G^c = \slc$  formulae are the same but all
entries are complex.
Let $\wt A^c$ be the complex subgroup
of all $2\times 2$ matrices of the form
\beq
\tilde h (r) =
\left ( \begin{array}{cc} r & 0 \\ 0 & {1\over r}\end{array} \right ),\ \ \
r \in \bC,\ \ \ r\not= 0\ .
\label{a.10}\endq
$\wt A$ is the subgroup of $\wt A^c$ in which $r >0$. Note that
$\wt A^c \cap \wt G = \wt A \cup -\wt A$ does not coincide with $\wt A$.
$Z_2 = \{1,\ -1\}$ is the common center of $\wt G^c$ and $\wt G$
and is contained in $\wt A^c $ (but not in $\wt A$).

$\wt G$ (resp. $\wt G^c$) operates on the real (resp. complex) 3-dimensional
Minkowski space $M_3$ (resp. $M_3^{(c)}$) by congruence: 
\begin{equation}
\label{similarity}
x \to X = \left(
\begin{array}{cc}
 x^0+x^1 &  x^2 \\
 x^2 & x^0-x^1 \\
\end{array}
\right),
\ \ \ \ 
 X' = \left(
\begin{array}{cc}
 {x^0}'+{x^1}' &  {x^2}' \\
 {x^2}' & {x^0}'-{x^1}' \\
\end{array}
\right) = \tilde g X \tilde g^T.
\end{equation}
The real (resp. complex) de Sitter manifold is mapped into itself by the above action, which is transitive {\em but not effective}.
In particular the subgroup $\wt A^c$ is seen to be the stability subgroup
of $(0,0,1)$ and the quotient $\wt G^c /\wt A^c$ can be identified
with the complex de Sitter space\footnote{Let $\wt H$ be a group,
$\wt K$ a subgroup of $\wt H$, and $Z \subset \wt K$ an invariant subgroup
of $\wt H$. Let $H= \wt H/Z$ and $K= \wt K/Z$.
Then $\wt H/\wt K \simeq H/K$. It follows that 
$\wt G^c /\wt A^c \simeq G^c/A^c$.}.
The real trace of the latter,
i.e. $\wt G/(\wt A\cup - \wt A)$,  can be identified to the real de Sitter space $dS_2$.
On the other hand $\wt G/\wt A$ can be identified to the two-sheeted
covering $\wt{dS_2}$ of $dS_2$. 
If $\tilde g \in \widetilde G^c$ and $r \not=0$,
\beq
\tilde g\tilde h(r)  =
\left ( \begin{array}{cc} ra & {b\over r} \\
rc & {d\over r}\end{array} \right )\ .
\label{a.30}\endq
In the real case when $\tilde g\in \wt G$ and $r > 0$ we can
take $r = (a^2 + c^2)^{-1/2}$ and get ${a'}^2 + {c'}^2 =1$ ($a$ and $c$ cannot be both equal to 0 because $ad-bc = 1$).
Thus every coset $\tilde g \wt A$  contains exactly one element
with this property;  it can be parametrized by using  the  Iwasawa decomposition\footnote{ The  Iwasawa parametrization of 
$\wt G = \slr$ is given by 
\begin{equation}
 \wtg=\tilde k(\zeta)\,\tilde n(\lambda)\,\tilde a(\chi) = \left(
\begin{array}{cc}
\cos\frac \zeta 2 & \sin\frac \zeta 2 \\
-\sin\frac \zeta 2  & \cos\frac \zeta 2 \\
\end{array}
\right) \left(
\begin{array}{cc}
 1 &  \lambda\\
 0 & 1 \\
\end{array}
\right) \left(
\begin{array}{cc}
e^{\frac \chi 2}& 0 \\
 0 &e^{-\frac \chi 2}\\
\end{array}
\right), \label{iwadec}
\end{equation}
where  $\lambda$ and $\chi$ are real and $\ttheta$ is a real number $\mod 4\pi$.
The parameters are related to $a,b,c$ and $d$ as follows:\begin{equation}\lambda = a b+c d, \ \ \ r=e^{\frac \chi 2} = {\sqrt{a^2+c^2}}, \ \ \ \cos\frac \zeta 2 =  \frac{a}{\sqrt{a^2+c^2}}, \ \ \  \sin\frac \zeta 2 = - \frac{c}{\sqrt{a^2+c^2}}.
\end{equation} }
of $\wt G$:
\begin{equation}
\tilde x (\lambda, \zeta)=\tilde k(\zeta)\,\tilde n(\lambda)= \left(
\begin{array}{cc}
 \cos \frac{\zeta}{2} &  \lambda  \cos  \frac{\zeta}{2} + \sin  \frac{\zeta}{2} \\
 -\sin  \frac{\zeta}{2}  & \cos  \frac{\zeta}{2} -\lambda  \sin  \frac{\zeta}{2}  \\
\end{array}
\right).
\end{equation}
The double covering $\widetilde {dS_2}$ can be thus represented as the following real algebraic manifold 
\beq
\widetilde {dS_2} = \wt G/\wt A \simeq \{(a,\ b,\ c,\ d) \in \bR^4\ :\ ad-bc = 1,\ \ a^2+c^2 =1\}\ 
\label{a.35}\endq
which is the intersection of two quadrics in $\bR^4$ and
can be verified to have no singular point.

$\slr$  acts on $\widetilde {dS_2}$ (i.e.  $\wt G /\wt A $) by left multiplication; the transformation rules formally coincide with the previous ones (see Eq. \ref{trasf1a} and Footnote \ref{fooy})   with the only difference that  the angular coordinates are now defined  $\mod \, 4\pi$. The action {\em is effective and transitive} and  $\slr$ can be interpreted as the relativity group of  $\widetilde {dS_2}$.


The Maurer-Cartan form 
provides   $\widetilde {dS_2}$ with a natural Lorentzian metric that may be constructed precisely as in the previous section
and one gets again Eq. (\ref{dscov}).  $\widetilde {dS_2}$ is a globally hyperbolic and geodesically complete spacetime.

 \subsection{$\slr$ and the three dimensional anti de Sitter universe} $\,$
 \vskip10pt
 
 There is an obvious bijection between the group $\slr$ and the three-dimensional anti de Sitter manifold 
\begin{equation}
AdS_{3}=\left\{ x \in {M}_{2,2}:\ {x^{0}}^{2}-{x^1}^{2}-{x^2}^{2}+{x^3}^2= 1 \right\} \label{ads}
\end{equation} 
 given by the identification
\begin{equation}
\left (\begin{array}{cc}a & b \\ c  & d\end{array} \right ) =
\left (\begin{array}{cc}x^0+x^1 & x^2+x^3 \\ x^2-x^3 & x^0-x^1
\end{array} \right ).
\label{slrads}
\end{equation}
It obviously follows that $\slr$ acts  transitively on $AdS_3$.
$\slr$ acting by left multiplication on the matrix in (\ref{slrads})
leaves its determinant unchanged so that $\slr$ is a subgroup of $SO(2,2)$.
This can also be seen by considering the mapping
\begin{equation}
\left(
\begin{array}{cc}
 a & b \\
 c & d \\
\end{array}
\right) \rightarrow  
\frac 1 2 \left(
\begin{array}{cccc}
 a+ d  &  a- d  & b+c & -b+c\\
 a- d   & a+ d   & b-c &-b-c\\
b+c&-b+c& a+ d  & a- d  \\
b-c & -b-c& a- d  &  a+ d  \\
\end{array}
\right).\label{33}
\end{equation}
The Iwasawa decomposition (\ref{iwadec}) and Eq. (\ref{33})  also provide
an  interesting {\em global} coordinate system for $AdS_3$.
An easy calculation explicitly shows that the Maurer-Cartan metric and the $AdS_3$ metric coincide :
\begin{eqnarray}
{\d}s^2 = \left.\left({\d x^0}^2-{\d x^1}^2-{\d x^2}^2+{\d x^3}^2\right)\right|_{AdS_3} 
= - \frac 12 \ {\rm Tr}\left( \tilde g^{-1} \, d\tilde g\,  \tilde g^{-1}\,  d\tilde g \right) = \\ = 
\frac{1}{4} \left(\text{d$\zeta^2 $} +2 \text{d$\zeta $} \text{d$\lambda $}-2  \lambda \text{d$\zeta $} \text{d$\chi $}
   -\text{d$\chi $}^2\right). \label{dsmet2a}
\end{eqnarray} 
(note that $\zeta$ is a timelike variable in the $AdS_3 $ manifold).
Setting $\chi=0$   we get a parametrization $x(\lambda,\ttheta,0)$ of $\widetilde{dS_2}$, represented here as the submanifold ${\widetilde{ \cal M}}$ of $AdS_3$:
 \begin{equation}
\widetilde {\cal M}=AdS_3\cap \{ (x^0+x^1)^2 + (x^2-x^3)^2=1\},  \ \ \ \ \ 
\wt x(\lambda,\ttheta, 0)= \left\{
\begin{array}{l}
x^0= \cos \frac{\zeta }{2}-\frac{1}{2} \lambda  \sin \frac{\zeta }{2} \\
x^1= \frac{1}{2} \lambda  \sin \frac{\zeta }{2} \\
x^2= \frac{1}{2} \lambda  \cos \frac{\zeta }{2} \\
x^3= \frac{1}{2} \lambda  \cos \frac{\zeta }{2}+\sin \frac{\zeta }{2} \\
\end{array} \label{coorads2}
\right.
\end{equation}
Of course $\wt{\cal M}$ is only invariant under the linear action of the image of the one-parameter subgroup $K$ of $\slr$, namely matrices of the form 
\begin{equation}
\left(
\begin{array}{rrrr}
 \cos \frac{\theta }{2}& 0 & 0 & -\sin  \frac{\theta }{2} \\
 0 & \cos \frac{\theta }{2}& \sin \frac{\theta }{2}& 0 \\
 0 & -\sin \frac{\theta }{2}& \cos  \frac{\theta }{2} & 0 \\
 \sin \frac{\theta }{2}& 0 & 0 & \cos \frac{\theta }{2} \\
\end{array}
\right)
\end{equation}
and we cannot simply take the restriction to {\cal M} of the $AdS_3 $ metric to get an $\slr$ invariant metric.

On the other hand $\slr$ acts on $AdS_3$ by congruence
\begin{equation}
\label{similarity2}
X = \left(
\begin{array}{cc}
 x^0+x^1 &  x^2 +x^3\\
 x^2 -x^3& x^0-x^1 \\
\end{array}
\right),
\ \ \ \ 
 X' = \left(
\begin{array}{cc}
 {x^0}'+{x^1}' &  {x^2}' +{x^3}' \\
 {x^2}' -{x^3}' & {x^0}'-{x^1}' \\
\end{array}
\right) = \tilde g X \tilde g^T.
\end{equation}
This provides the  map
\begin{equation}
\left(
\begin{array}{cc}
 a & b \\
 c & d \\
\end{array}
\right) \rightarrow  
 \frac{1}{2}\left(
\begin{array}{cccc}
 \left(a^2+b^2+c^2+d^2\right) &  \left(a^2-b^2+c^2-d^2\right) &  (2 a b+2
   c d) & 0 \\
  \left(a^2+b^2-c^2-d^2\right) &  \left(a^2-b^2-c^2+d^2\right) &(2 a b-2
   c d) & 0 \\
2 a c+2 b d & 2 a c-2 b d & 2 b c+2 a d & 0 \\
 0 & 0 & 0 & 2 \\
\end{array}
\right) \label{11}
\end{equation}
Note that $\tilde g$ and $-\tilde g$ are mapped into the same element
of $SO(2,2)$; indeed Eq. (\ref{11}) is a covering projection of $\slr$
onto a $SO_0(1,2)$ subgroup of $SO(2,2)$.  Using again the Iwasawa parameters
we  get a coordinate system for $dS_2$  represented here as
a submanifold ${\cal M}$ of $AdS_3$
 \begin{equation}
{\cal M}=AdS_3\cap \{ x^3=\sqrt 2\},  \ \ \ \ \ 
x(\lambda,\ttheta, 0)= \left \{
\begin{array}{c}
 \lambda  \\
 \lambda  \cos (\zeta )+\sin (\zeta ) \\
 \cos (\zeta )-\lambda  \sin (\zeta ) \\
 \sqrt{2} \\
\end{array}
\right .
\end{equation}
Of  course the restriction of the $AdS$ metric to ${\cal M}$ gives back Eq. (\ref{dsmet}).

\subsection{Complexification}
The algebraic manifold (\ref{a.35}) 
can be complexified, i.e. we can define
\beq
\VV = \{(a,\ b,\ c,\ d) \in \bC^4\ :\ ad-bc = 1,\ \ a^2+c^2 =1\}\ .
\label{a.40}\endq
Let us again consider eq. (\ref{a.30}) but now with complex $\tilde g$ and
$r$. If $\tilde g$ is such that $a^2 + c^2 \not= 0$,
we can choose $r = \pm (a^2 + c^2)^{-1/2}$
and thus the coset $\tilde g\widetilde A^c$ contains two distinct (opposite)
elements of $\VV$. If $p$, $p'$ are points of $\VV$
such that $p \not= \pm p'$ there is no $r \not= 0$ such that $p'= p\tilde h(r)$ 
hence $p$ and $p'$ belong to different elements of $\wt G^c/\wt A^c$.
Conversely two opposite points $p$ and $-p$ of the manifold $\VV$ belong to the
same coset $p\wt A^c$ i.e. determine a unique element
of $\wt G^c/\wt A^c$. On the other hand if $\tilde g$ is such that $a^2+c^2 = 0$, all
elements of the coset $\tilde g\wt A^c$ have the same property and none belongs to
$\VV$. 

The cosets $\wt g\wt A^c$ such that $a^2+c^2 = 0$ can be identified to
certain points of the complex de Sitter space as follows.
Let $g = (a,\ b,\ c,\ d) \in \wt G^c$. Then 
$x = g(0,\ 0,\ 1)$ is given by
\beq
\left ( \begin{array}{cc} x^0 + x^1 & x^2 \\ x^2 & x^0 - x^1
\end{array} \right ) = 
\left ( \begin{array}{cc} a & b\\ c & d \end{array} \right )
\left ( \begin{array}{cc}0 & 1\\ 1 & 0 \end{array} \right )
\left ( \begin{array}{cc} a & c\\ b & d \end{array} \right ) =
\left ( \begin{array}{cc} 2ab & ad + bc\\ ad + bc & 2cd \end{array} \right )\ 
\label{a.45}\endq
(the fact that the determinant of the lhs is equal to $-1$ expresses
$x \in dS_2^{(c)}$).
Since $a$ and $c$ cannot be both 0 we may suppose that $c \not= 0$; we get
\beq
{a\over c} = {x^0+x^1\over x^2-1} = {x^2+1\over x^0 - x^1}\ .
\label{a.47}\endq
The condition $a^2 + c^2 = 0$ is equivalent to $a = \pm ic$
(thus excluding $c=0$) and implies
\beq
x^0+x^1 = \pm i(x^2-1) \Longrightarrow x^0 - x^1 = \mp i (x^2+1)\ .
\label{a.48}\endq
Conversely one of these conditions implies $a^2+c^2 = 0$. Therefore
\beq
a^2+c^2 = 0 \Longleftrightarrow (x^0+x^1)^2 + (x^2-1)^2 = 0
\Longleftrightarrow (x^0-x^1)^2 + (x^2+1)^2 = 0\ .
\label{a.49}\endq
It follows that $\VV$ projects onto $dS_2^{(c)}$ with the exception of the
manifold $\NN$ defined by the above equations. In particular the points
$(\pm i,\ 0,\ 0)$ belong to $\NN$. Of course the manifold $\NN$ is not
invariant under the action of $\wt G$ or $\wt G^c$.

As a parenthesis let us take advantage of the preceding calculations to verify
that $\wt G^c$ acts transitively on $dS_2^{(c)}$, i.e. given
$(x^0,\ x^1,\ x^2) \in \bC^3$ satisfying ${x^0}^2 - {x^1}^2 - {x^2}^2 = -1$,
there is a $g = (a,\ b,\ c,\ d)$ such that $ad -bc = 1$ and eq. (\ref{a.45})
holds. %

There are other complex manifolds that contain $\wt{dS_2}$ as a real form.
For example let us represent $\wt{dS_2}$ as a cylinder $\bR\times S^1$
as follows: a point is associated to a pair $(x^0,\ \theta)$ where
$x^0 \in \bR$ and $\theta \in \bR/4\pi\bZ$. It projects on the point
\beq
\left ( \begin{array}{c}x^0\\ \sqrt{{x^0}^2 +1} \cos \theta\\
\sqrt{{x^0}^2 +1} \sin \theta \end{array}\right ) \in dS_2\ .
\label{a.60}\endq
The natural complexification of this is $\Sigma \times \bC/4\pi\bZ$,
where $\Sigma$ is the Riemann surface of $z \mapsto \sqrt{z^2+1}$,
a two-sheeted covering of $\bC \setminus \{i,\ -i\}$.
This complex manifold projects onto $dS_2^{(c)}$ with the exception
of the points such that $({x^0}^2 +1) = 0$, or equivalently
$({x^1}^2 + {x^2}^2) = 0$.

Another example will be used to discuss the analyticity of certain two-point
functions in Sect. \ref{conver}. The real space $dS_2$ (resp. $\wt {dS_2}$)
can be diffeomorphically mapped onto a slice of the cylinder
$\bR\times \bR/2\pi \bZ$ (resp. $\bR\times \bR/4\pi \bZ$) as follows (using the
notation of (\ref{parglo})):
\begin{align}
&x(t, \theta) \mapsto (s, \theta),\cr
&\sh t = \tg s,\ \ s = 
{1\over 2i} \log \left({1+i\sh t \over 1-i\sh t}\right )
= {1\over 2i} \log \left({1+ix^0\over 1-ix^0}\right )  \ ,
\label{a.70}\end{align}
As $t$ varies in $\bR$, $s$ varies in $(-\pi/2,\ \pi/2)$,
and as a consequence
\beq
x^0 = \tg s,\ \ \ 
\ch t = {1\over \cos s},\ \ \ \sin s= \th t,\ \ \
t = \haf \log \left ({1+\sin s \over 1-\sin s } \right )\ .
\label{a.75}\endq
This map is conformal:
\beq
dt^2 - \ch^2 t  \, d\theta^2 = {ds^2 - d\theta^2 \over \cos^ 2\, s}.
\label{a.76}\endq
The cylinder $\bR\times \bR/2\pi \bZ$ (resp. $\bR\times \bR/4\pi \bZ$)
can be complexified as $\bC\times \bC/2\pi \bZ$
(resp. $\bC\times \bC/4\pi \bZ$). However the map (\ref{a.70}) becomes
singular when $x^0 = \pm i$, so that these complex cylinders
project onto $dS_2^c$ with the exception of the points where $x^0 = \pm i$.
Note that the complexified map, when restricted to the 'Euclidian'
2-sphere (minus the poles $x^0 = \pm i$), is the Mercator projection
of that sphere.

The preceding examples are all of the following type. $\wt {dS_2}$ is imbedded
in a connected
complex manifold $\wt \MM$ equipped with a complex conjugation $*$, and
$\wt {dS_2} \subset \wt \MM^{(r)} = \{x \in \wt \MM\ :\ x=x^* \}$.
The projection $\wt\pr$ of  $\wt {dS_2}$ onto $dS_2 \subset dS_2^c$
can be analytically continued to an analytic local 
homeomorphism of $\wt \MM$ onto an open subset $\MM$ of $dS_2^c$ so that
$\wt \MM$ together with $\wt\pr$ determine a covering space of $\MM$.
But $\MM$ can never
be the whole $dS_2^c$. Indeed the latter being simply connected, $\wt\pr$
would be a homeomorphism of $\wt \MM$ onto $dS_2^c$. This is not possible
since the restriction of $\wt\pr$ to $\wt {dS_2}$ is 2 to 1.
$\MM$ cannot be invariant under $G^c$ since $G^c$ acts transitively
on $dS_2^c$, and this would imply $\MM = dS_2^c$. However the examples
mentioned above are invariant under the subgroup of rotations.
Note that in the case of the third example mentioned above,
$\wt \MM$ is the cylinder $\bC\times \bC/4\pi\bZ$ minus the points
where $s \in {\pi\over 2}+ \pi\bZ$, and the projection
$\wt \pr\ :\ \wt \MM \rightarrow dS_2^c$ is given by the formulae
(\ref{a.75}), i.e.
$(s,\ \theta) \rightarrow (x^0 = \tg s,\ \theta\ \mod 2\pi)$.


\section{More about the covering projection}
We saw that he group  $\widetilde G= \slr$  acts on the covering space
$\widetilde{dS_2}$ as a group of {\em spacetime transformations} by left
multiplication   ${\tilde{x}}\to \wtg{\tilde{x}}$  and  acts on $dS_2$ by congruence   (\ref{similarity}). We denote both actions by the shortcut $(\cdot )\to  \wtg (\cdot) $. They commute with the covering projection (\ref{proju}):
\beq
\pr(\wtg\wtx) = \wtg\, \pr(\wtx)\ \ \ \forall \wtg\in \widetilde G,\ \ \forall \wtx\in \widetilde {dS_{2}}\ .
\label{m.30}\endq

On the de Sitter manifold $dS_2$ the antipodal map $x\mapsto -x$ is expressed 
in the coordinates $(t,  \theta)$ by 
$x(t,  \theta)\mapsto -x(t,  \theta) = x(-t,  \theta+\pi).$
Let $\tau_1$ be the operation with the same expression on the covering manifold $\widetilde {dS_{2}}$, i.e.
\beq
\tau_1 \tilde x (t,  \theta) = \tilde x(-t,  \theta+\pi),\ \ \  \tilde x \in \widetilde {dS_{2}}
\label{m.35.a}\endq
and let  
\beq
\tau_2\tilde x(t, \ \theta) = \tilde x(-t,  \theta -\pi),\ \ \
 \tau\tilde x(t, \ \theta) = \tilde x(t,  \theta+2\pi),\ \ \
 \tilde x \in \widetilde {dS_{2}}, \ \ \ \tau = \tau_1^2 = \tau_2^2,\ \ \
\tau_1\tau_2 = 1\ .
\label{m.36}\endq
 Obviously $\tau_1$ is a diffeomorphism of the covering manifold; we have 
\beq
\pr(\tau_1 \wtx) = - \pr(\wtx),\ \ \ \pr(\tau \wtx) =  \pr(\wtx)\, .
\label{m.37}\endq
\ifundefined{infinitesimals}
The following lemma proves that  $\tau_1$ commutes
with the action of the group $\tilde G$.

\begin{lemma}
Let $\vhi$ (resp. $\wt \vhi$) be a continuous map of $dS_2$ (resp.  $\wt {dS_{2}}$)
into itself such that 
$\pr \wt\vhi (\tilde x) = \vhi (\pr \wtx)$  for all $\wtx\in\wt {dS_{2}}$. Suppose that, for every $y\in dS_2$
and every $g\in  G $, $\vhi(g y) = g\vhi(y)$. Then, for every
$\wtx\in \wt {dS_{2}}$  and every $\wtg\in \wt G$, $\wt\vhi(g \wtx) = \wtg\wt\vhi(\wtx)$.
\end{lemma}

{\bf Proof.}
Let $\wt x \in \wt{dS_{2}}$. 
Let $B$ be the set of all $g \in \wt G $ such
that $\wt\vhi(g \wt x) = g\wt\vhi(\wt x)$. $B$ contains the identity
and is obviously closed. Let $g \in B$ and let $V$ be an open neighborhood
of $\wt\vhi(g \wt x) = g\wt\vhi(\wt x)$ such that $\pr$ is a diffeomorphism
of $V$ onto $\pr\, V$. Let $W$ be an open neighborhood of $g$ in $\wt G$ such
that $h\wt\vhi(\wt x)\in V$ and $\wt\vhi( h \wt x) \in V$
for all $h$ in $W$. For any $h\in\wt G$, 
\beq
\pr h\wt\vhi(\wt x) = h \pr \wt\vhi(\wtx) = h \vhi(\pr \wt x) = \vhi(\pr h\wt x) =
\pr\wt\vhi( h \wt x)\ .
\label{m.38}\endq
Since $\pr$
is a diffeomorphism on $V$, $h \in W$ implies that
$\wt\vhi(h \wt x) = h\wt\vhi(\wt x)$, i.e.
$h\in B$, i.e.  $W\subset B$. Thus $B$ is open and must coincide with $\wt G$.
If we take $\wt\vhi(\wt x) = \tau_1 \wt x$ and $\vhi( x) = -x$ we obtain
$g\tau_1 \wt x = \tau_1 g \wt x$ for all $g\in \wt G$.
\else
To prove that $\tau_1$ commutes with the action of the group $\tilde G$,
it suffices to prove that it commutes with the elements of a basis of the linear
differential operators associated with the action of the Lie algebra of
$\wt G$ on $\wt{dS_2}$, e.g.
the operators defined on $\CC^\infty (\wt{dS_2})$ by
\begin{align}
&\MM_j f(x) = {d\over d\tau}f(e^{\tau M_j}x) \Big |_{\tau=0}\ ,\ \ \
j = 0,\ 1,\ 2,\cr
&M_0 = \haf \left ( \begin{array}{cc} 0 & -1 \\ 1 & 0 \end{array}\right )\ ,\ \ 
M_1 = \haf \left ( \begin{array}{cc} 0 & 1 \\ 1 & 0 \end{array}\right )\ ,\ \ 
M_2 = \haf 
\left ( \begin{array}{cc} 1 & 0 \\ 0 & -1 \end{array}\right )\ .
\label{ft.10}\end{align}
In the coordinates $(t,\ \theta)$ defined in (\ref{parglo}),
\begin{align}
& \MM_0 = -{\partial \over \partial \theta}\ ,
\label{m.40.1}\\
& \MM_1 = \cos(\theta){\partial \over \partial t}
- \sin(\theta) \th(t) {\partial \over \partial \theta}\ ,
\label{m.40.2}\\
& \MM_2 = \sin(\theta){\partial \over \partial t}
+\cos(\theta) \th(t) {\partial \over \partial \theta}\ .
\label{m.40.3}\end{align}
It is easy to verify that
\beq
[\MM_1\,,\ \MM_2] = -\MM_0\,,\ \ \ 
[\MM_1\,,\ \MM_0] = -\MM_2\,,\ \ \ 
[\MM_2\,,\ \MM_0] = \MM_1\ .
\label{v.33}\endq
Consider now a smooth function $f(t,\ \theta)$ (not necessarily
periodic in $\theta$) and let $f_t$ and $f_\theta$ denote its partial
derivatives. We need to verify that 
\beq
\MM_j(f\circ \tau_1) = (\MM_j f)\circ \tau_1\ ,\ \ \ j= 0,\ 1,\ 2\ ,
\label{m.41}\endq
This is obvious in the case of $\MM_0$. In the case of $\MM_1$
\begin{align}
& \MM_1(f\circ \tau_1)(t,\ \theta) = 
-\cos(\theta)f_t(-t,\ \theta+\pi)
- \sin(\theta) \th(t)f_\theta(-t,\ \theta+\pi)\ ,\cr
&(\MM_1 f)\circ \tau_1(t,\ \theta) =
\cos(\theta+\pi)f_t(-t,\ \theta+\pi)
- \sin(\theta+\pi) \th(-t)f_\theta(-t,\ \theta+\pi)\ ,
\label{m.42}\end{align}
and these two expressions are equal. The case of $\MM_2$ is similar.
\fi
It follows that $\tau$ and $\tau_2$ also commute with the action of $\wt G$.

\ifundefined{infinitesimals}{}
\else
We also note that a necessary and sufficient condition
for a smooth function or a distribution $f$ on
$\wt{dS_2}$ to be invariant under $\wt G$ is that $\MM_jf = 0$ for all
$j= 0,\ 1,\ 2$, but in fact it suffices that this hold for $j = 0$ and $j= 1$
(or for $j = 0$ and $j= 2$)
because of (\ref{v.33}). Similarly a smooth function or distribution $f$ on
$\wt{dS_2} \times \wt{dS_2}$ is invariant if and only if
$(\MM_{j,\wtx}+\MM_{j,\wtx'}) f(\wtx,\ \wtx') =0$ for $j = 0,\ 1$
(or for $j = 0$ and $j= 2$).
\fi

\section{Quantum field theory on the 2-dim dS universe vs its double covering}

In the spherical coordinate system (both on $dS_2$ and on $\widetilde{dS_2}$) the de Sitter Klein-Gordon equation  takes the form
\begin{equation}
\Box \phi - \lambda(\lambda+1) \phi=
\frac{1}{\cosh t} \ \partial_t (\cosh t \ \partial_t \phi)
- \frac{1}{\cosh^2 t} \ \partial^2_{\theta} \phi - \lambda(\lambda+1)\phi =0 .
\label{s.16}
\end{equation}
The parameter $\lambda$ and the squared mass 
\beq
m^2_\lambda = - \lambda(\lambda+1) \label{masssquared}
\eeq are  complex numbers;  
$m^2_\lambda$ is real and positive in the following special cases:
\begin{eqnarray}
{\rm either\ \ }&&   \lambda = -\frac {1}2 + i \rho, 
\ \ \ \Im \rho = 0,\ \ \ m = \sqrt{\frac 1 4+ \rho^2 }\geq \frac{1}{2}\, ,
\label{principal}
\\
{\rm or\ \ }&&  \Im \lambda =0, 
\ \ \ -1<\Re \lambda <0, \ \ \ 0<m< \frac{1}{2}.
\label{reallambda}
\end{eqnarray}
Let us  introduce the complex variable $z= i \sinh t$,
so that $1-z^2 = \cosh^2 t$,
and separate the variables by posing
$\
\phi =  f(z) e^{{i  l \theta}}
\label{s.25} $. 
 Eq. (\ref{s.16}) implies that $f$  has to solve the Legendre differential equation:
\beq
(1-z^2) f''(z) -2zf'(z) + \lambda(\lambda+1) f(z) -{l^2 \over (1-z^2)}f(z) = 0.
\label{s.19}\endq
The difference between $dS_2$ and its covering $\widetilde{dS_2}$ is that in the first case  $l$ is an integer number while in the second case $2l$ is integer. Enlarging the set of possible values of $l$ in this way will cause many unexpected new features. We will describe some of them below.

Two linearly 
independent\footnote{The following formulae are useful to compute the various Wronskians
$
{\cal W}\{w_1,w_2\}=w_1(z)w^{\prime}_2(z)-w_2(z)w^{\prime}_1(z)
$
needed:
\begin{eqnarray}
\P^\mu_\nu(0)&=& \frac{2^{\mu +1} \sin \left(\frac{1}{2} \pi  (\mu +\nu )\right) \Gamma \left(\frac{1}{2}
   (\mu +\nu +2)\right)}{\sqrt{\pi } \Gamma \left(\frac{1}{2} (-\mu +\nu +1)\right)}, \label{po} \\
{ {\P'}^\mu_\nu(0)} &=& \frac{2^{\mu } \cos \left(\frac{1}{2} \pi  (\mu +\nu )\right) \Gamma \left(\frac{1}{2}
  (\mu +\nu +1)\right)}{\sqrt{\pi } \Gamma \left(\frac{1}{2} (-\mu +\nu +2)\right)}.
\end{eqnarray}
where $\P'(z) = \frac {d\P}{dz}$. We get
\begin{eqnarray}
{\cal W}
\{\P^\mu_\nu(i \sinh t) , 
\P^\mu_\nu(-i \sinh t) \} &=&\frac{2}{\Gamma (-\mu -\nu ) \Gamma (-\mu +\nu +1)}\,({\cosh t})^{-2}
\cr
{\cal W}
\{\P^\mu_\nu(i \sinh t) , 
\P^{-\mu}_\nu(i \sinh t) \}&=& -\frac {2  \sin(\pi\mu)}\pi   ({\cosh t})^{-2}
\cr
{\cal W}
\{\P^\mu_\nu(-i \sinh t) , 
\P^{-\mu}_\nu(-i \sinh t) \}&=& \frac {2  \sin(\pi\mu)}\pi   ({\cosh t})^{-2}
\cr
{\cal W}
\{\P^\mu_\nu(i \sinh t) , 
\P^{-\mu}_\nu(-i \sinh t)  \} &=& -\frac{2 \sin (\pi  \nu )}{\pi }  ({\cosh t})^{-2}
\end{eqnarray}}
 solutions of the above equation are the  Ferrers functions (also called
``Legendre functions on the cut'' \cite{HTF1}) $\P_\nu^\mu(z)$ and $\Q_\nu^\mu(z)$,
where 
$$
\nu = \lambda , \, \mu = -l . 
$$
 $\P_\nu^\mu(z)$ and $\Q_\nu^\mu(z)$ are holomorphic in the
cut-plane
\begin{equation}
\Delta_2 =
{\bf C}\setminus (-\infty-1] \cup [1,\infty) .
\label{t.22}\end{equation}
and satisfy the  reality conditions
\begin{equation}
\ovl{\P_\nu^\mu(z) }= {\P_{\ovl{\nu}}^{\ovl{\mu}}(\bar z)}, \ \ \
\ovl{\Q_\nu^\mu(z)} = {\Q_{\ovl{\nu}}^{\ovl{\mu}}(\bar z)} \label{reality}
\end{equation}
 for all $z\in\Delta_2$.  $\P_\nu^\mu(z)$ respects the symmetry (\ref{masssquared})  of the mass squared: for all $z\in\Delta_2$ it  satisfies the identity \cite{HTF1}
\beq
\P_\lambda^{-l}(z) =\P_{-\lambda-1}^{-l}(z).
\label{t.22.1}\endq
If $\lambda-l$ and $\lambda+l-1$ are not non-negative integers,  
$\P_\lambda^{-l}(z)$ and 
$\P_\lambda^{-l}(-z)$
also constitute two linearly independent solutions of Eq. (\ref{s.19}).
In this case the general solution has
the form
\begin{equation}
\phi^{}_{  l}(t,\x) =  
[a_{ l}\, \P^{- l }_{\lambda}(i\sh t)
+b_{  l}\, \P^{- l }_{\lambda}(-i\sh t)]\ e^{{il \theta}} 
.
\label{modesab0}
\end{equation}
In the following  we will restrict our attention to values of $\lambda$
such that 
\beq
-{1\over 2} < \Re \lambda < 0\ .
\label{t.23}\endq
In partiular we do not consider here tachyonic fields.

\section{Canonical  commutation relations.}

Let us focus on the  modes 
\begin{eqnarray}
\phi_{ l}(t,\x) &= & 
[a_l\P^{- l }_{\lambda}(i\sh t) +b_l  \P^{- l }_{\lambda}(-i\sh t)]e^{{il \theta}} \cr
\phi^*_{ l}(t,\x)& =  &
[a^*_l\P^{- l }_{\lambda}(-i\sh t) +b^*_l  \P^{- l }_{\lambda}(i\sh t)]e^{{-il \theta}} 
\label{modesab01}
\end{eqnarray}
where either $\lambda = -1/2 +i\nu$ or $\lambda$ real. The  KG product is defined as usual \cite{Birrell}:
\begin{equation}
(f,g)_{KG} = 
i \int_\Sigma (f ^*\partial_\mu g -g\, \partial_\mu f ^* )  d\Sigma^\mu(x) = i \int_\Sigma  (f ^*\partial_t g -g\, \partial_t f ^* )  d\theta
\label{kgprod}\end{equation}
On $dS_2$ the integral is over the interval $\Sigma= [0,2\pi]$ and $l$ is integer. When we consider fields on the covering manifold $\widetilde{dS_2}$ the  integral is over the interval $\Sigma = [0,4\pi]$ and $2l$ is integer.

The first condition imposed by the canonical quantization procedure is the orthogonality  $(\phi^*_{l}, \phi^{}_{ l'})_{KG} =0$ of the modes;
%
%
%
it gives rise to  the following conditions on the coefficients:
\begin{eqnarray}
&& a_l  b_{-l}  -b_l  a_{-l}  =0 \ \ \ \makebox{for $l\in{\Bbb Z}$  (i.e. on both $dS_2$ and $\widetilde{dS_2}$), } \label{kgpds}\\ 
&& a_l  a_{-l}  -b_l  b_{-l} = c_l\sin (\pi  \lambda ), \ \ a_l  b_{-l}  -b_l  a_{-l}  =c_l \sin(\pi l) \ \ \makebox{for $l\in\frac 12 +{\Bbb Z}$ (only on $\widetilde{dS_2}$). }
\end{eqnarray}
 The constants $c_l$ are unrestricted by the above conditions, which are summarized as follows:
\begin{eqnarray}
a_{-l}=c_l ({{a_l} \sin (\pi  \lambda )+{b_l} \sin (\pi  l)}),  \ \ \ 
b_{-l}=c_l ({{b_l} \sin (\pi  \lambda )+{a_l} \sin (\pi  l)}). \label{kkgg2}
\end{eqnarray}
The normalization condition is given by 
\begin{eqnarray}
(\phi^{}_{  l},\phi^{}_{ l' })_{KG}
= \frac {2 k\pi }{\gamma_l}\,(|a_{ l}|^2 - |b_{ l}|^2) \, \delta_{ll'}  = \frac{1}{N_l} \delta_{ll'} 
\label{kkgg1}
\end{eqnarray}
where $k=1$ for $dS_2$ and $k=2$ for $\widetilde{dS_2}$ and 
\begin{equation}
\gamma_l = \frac 12 {\Gamma(l-\lambda) \Gamma(1 +\lambda + l ) } \label{gammaal}
\end{equation}
so that 
\begin{equation}
N_l =  \frac {\gamma_l }{2 k\pi (|a_{ l}|^2 - |b_{ l}|^2)} = 1. \label{kkgg12}
\end{equation}
As a function of $l $,  the product $ \gamma_l $ is always positive for $\lambda = -\frac 12 +i \nu$. 
When $-1 <\lambda <0$ it takes negative values for negative half integer $l$'s.

The commutator finally takes the following form:
\begin{eqnarray} 
&& C(t,\theta,t',\theta')=\sum_{kl \in {\Bbb Z}}  N_l[\phi_{  l}(t,\x)
\phi^*_{  l}(t',\x') - \phi_{  l}(t',\x')
\phi^*_{  l}(t,\x)] = \label{ccrcov}
\cr &&
\cr
&&= \sum_{kl \in {\Bbb Z}}  N_l(|a_l|^2-|b_l|^2) [\P^{- l }_{\lambda}(i\sh t) \P^{- l }_{\lambda}(-i\sh t')- \P^{- l }_{\lambda}(-i\sh t) \P^{- l }_{\lambda}(i\sh t')] \cos{(l \theta-l \theta')} 
\cr
&& + \sum_{kl \in {\Bbb Z}}  i N_l(|a_l|^2+|b_l|^2) [\P^{- l }_{\lambda}(i\sh t) \P^{- l }_{\lambda}(-i\sh t')+ \P^{- l }_{\lambda}(-i\sh t) \P^{- l }_{\lambda}(i\sh t')] \sin{(l \theta-l \theta')} 
\cr
&& + \sum_{kl \in {\Bbb Z}} [2  i N_l a_l b_l^* \P^{- l }_{\lambda}(i\sh t) \P^{- l }_{\lambda}(i\sh t')+2  i N_l a^*_l b_l \P^{- l }_{\lambda}(-i\sh t) \P^{- l }_{\lambda}(-i\sh t')] \sin{(l \theta-l \theta')} .
\cr
&&
\end{eqnarray}
where, again, $k=1$ for $dS_2$ and $k=2$ for $\widetilde{dS_2}$ . We left in this expression explicitly indicated $N_l$ as a function of $a_l$ and $b_l$, as in Eq. (\ref{kkgg12}), even though the normalization condition imposes $N_l=1$. This allows to verify the locality property of the above expression more easily. Let us  indeed prove that the  equal time commutator 
\begin{eqnarray} 
C(0,\theta,0,\theta')=  2 i \sum_{kl \in {\Bbb Z}}   \frac {\gamma_l(|a_l+b_l|^2)}{2k \pi \,(|a_{ l}|^2 - |b_{ l}|^2)}
 [ \P^{- l }_{\lambda}(0)]^2 \sin{(l \theta-l \theta') } 
\end{eqnarray}
vanishes. The terms contributing to $C(0,\theta,0,\theta')$ are  the ones antisymmetric in the exchange of $\theta$ and $\theta'$ (the second and third line in Eq. (\ref{ccrcov})).
By using Eq. (\ref{kkgg2}) we have that 
\beq
{ \frac{|a_l+b_l|^2}{|a_{l}|^2-|b_{l}|^2 } }=
 {\cot  \left(\frac 12 \pi(l+  \lambda)\right ) }{\tan\left(\frac 12 \pi  (\lambda - l)\right) } {\frac{|a_{-l}+b_{-l} |^2}{|a_{-l}|^2-|b_{-l}|^2} }
\eeq 
In deriving the above identity we took in to account the hypothesis $\lambda = -1/2 +i\nu$ or $\lambda$ real, which implies that $\sin\pi \lambda$ is a real number. 

On the other hand  formula (\ref{po}) gives 
\begin{equation}
\frac{\gamma_l  \P^{- l }_{\lambda}(0)^2}{\gamma_{-l}  \P^{ l }_{\lambda}(0)^2} = \tan \left(\frac{1}{2} \pi  (\lambda +l)\right) \cot \left(\frac{1}{2} \pi 
   (\lambda-l )\right).\end{equation}
Therefore the coefficients of $\sin [l (\theta - \theta')]$ and of $\sin[-l(\theta - \theta')]$ are equal and the equal time commutator $C(0,\theta,0,\theta')$ vanishes. 

Let us verify now the CCR's:
\begin{eqnarray} 
&& \partial_{t'}C(t,\theta,t,\theta')|_{t=t'=0}=  -2 i  \sum_{kl \in {\Bbb Z}}   \frac {\gamma_l}{2 k \pi}
 [ \P^{- l }_{\lambda}(0){ \P'}^{- l }_{\lambda}(0)] \cos{(l \theta-l \theta')} +\cr 
 && + 2 i  \sum_{kl \in {\Bbb Z}} \sum  \frac {\gamma_l}{2k\pi (|a_{l}|^2-|b_{l}|^2) }(a_l b_l^*-a^*_l b_l)
 [ \P^{- l }_{\lambda}(0){ \P'}^{- l }_{\lambda}(0)] \sin{(l \theta-l \theta')} =\cr 
&& =  i  \sum_{kl \in {\Bbb Z}}   \frac {1 }{2k\pi} \cos{(l \theta-l \theta')} = i  \delta(\theta-\theta'\ \mod 2k\pi) \label{ccr}
\end{eqnarray}
where we used again Eq. (\ref{kkgg2}) 
As a byproduct we deduce that the second and third line in Eq. (\ref{ccrcov}) vanish identically and the covariant commutator may be re-expressed as follows: 
\begin{eqnarray} 
&& C(t,\theta,t',\theta') =
\sum_{kl \in {\Bbb Z}} \frac {\gamma_l}{2k\pi } [\P^{- l }_{\lambda}(i\sh t) \P^{- l }_{\lambda}(-i\sh t') - \P^{- l }_{\lambda}(-i\sh t) \P^{- l }_{\lambda}(i\sh t')] \cos{(l \theta-l \theta')}\  \cr && =
\sum_{kl \in {\Bbb Z}} \frac {\gamma_l }{2k \pi } [\P^{- l }_{\lambda}(i\sh t) \P^{- l }_{\lambda}(-i\sh t') - \P^{- l }_{\lambda}(-i\sh t) \P^{- l }_{\lambda}(i\sh t')] \exp{(i l \theta-i l \theta')}.  \label{coeff2} 
\end{eqnarray}
The second step follows from the symmetry of the generic term of the series at the right hand side of Eq. (\ref{coeff2}) 
under the change $l\to -l$.

\section{$\slr$-invariance of the commutator}
\label{seccomminv}
While the $\slr$ invariance of the commutator is a priori guaranteed by the vanishing of the equal time commutator and by the CCR's (\ref{ccr}),  it is instructive for what follows to give a direct proof based on the recurrence relations satisfied by the Legendre functions on the cut \cite{HTF1}. This will  prepare the task of  Section \ref{proof}, where the question of finding the more general invariant two-point function will be addressed.

To this aim, let us start by considering the first term at the RHS of Eq. (\ref{coeff2}) : 
\begin{eqnarray}
&& \twofcn_0 (\tilde x,\tilde x') = \frac 1 {4\pi}\sum_{2l \in {\Bbb Z}}   \gamma_l
\left[ \P^{-l}_{\lambda}(i\sinh t)
\P^{-l}_{\lambda}(-i\sh t') \right]  \exp(il(\theta-\theta')).\label{tpabh0}
\end{eqnarray}
This kernel is non local\footnote{The word local here and everywhere refers to {\em local commutativity}. We stress again that locality on the de Sitter manifold and on its covering are two distinct notions.}  but it turns out be $\slr$-invariant. 
The proof of this statement amounts to checking that the following infinitesimal condition holds:
\beq
\delta \twofcn= \sin \theta \, \partial_t \twofcn +\tanh t\, \cos \theta\,  \partial_\theta \twofcn+\sin \theta' \, \partial_{t'} \twofcn +\tanh t'\, \cos \theta'\,  
\partial_{\theta'} \twofcn = 0  \label{conddd}
\eeq
namely
\begin{eqnarray} \delta \twofcn_0 = 0 =  \sum_{l}  \gamma_l  \left[ i \sin \theta \cosh t \,  {{\P'}^{-l}_{\lambda}}(z)+ i l  \tanh t\, \cos \theta\,  \P^{-l}_{\lambda}(z) \right] \P^{-l}_{\lambda}(-z')\,e^{il(\theta-\theta')}  \cr -\sum_{l}  \gamma_l \P^{-l}_{\lambda}(z) \left[ i \sin \theta' \cosh t' \,  {{\P'}^{-l}_{\lambda}}(-z')+ i l  \tanh t'\, \cos \theta'\,  \P^{-l}_{\lambda}(-z') \right] e^{il(\theta-\theta')} .
\end{eqnarray}
Singling out the Fourier coefficient of $\exp(i l \theta)$, the above condition translates into the following requirement:
\begin{eqnarray}
&&\gamma_{\m-1} \P^{1-\m}_{\lambda}(-z' )  e^{i\theta'}\ \left[  \cosh t \,  {{\P'}^{1-\m}_{\lambda}}(z)+ i (\m-1)  \tanh t\, \,  \P^{1-\m}_{\lambda}(z) \right]+\cr 
 &+& \gamma_{\m+1} \ \P^{-1-\m}_{\lambda}(-z' )  e^{-i\theta'}\left[ - \cosh t \,  {{\P'}^{-1-\m}_{\lambda}}(z)+ i (\m+1)  \tanh t\, \,  \P^{-1-\m}_{\lambda}(z) \right] +\cr
  &-& 2\gamma_\m \P^{-\m}_{\lambda}(z) \left[ i \sin \theta' \cosh t' \,  {{\P'}^{-\m}_{\lambda}}(-z')+ i \m  \tanh t'\, \cos \theta'\,  \P^{-\m}_{\lambda}(-z') \right]=0. \label{condf}
\end{eqnarray}
%
This expression may be simplified by using the following  crucial identities:
\begin{eqnarray} &&  \cosh t \,  {{\P'}^{1-\m}_{\lambda}}(z)+ i (\m-1)  \tanh t\, \,  \P^{1-\m}_{\lambda} (z) \cr
&=&    (\lambda -\m+1) (\lambda +\m) \left[ - \cosh t \,  {{\P'}^{-1-\m}_{\lambda}}(z)+ i (\m+1)  \tanh t\, \,  \P^{-1-\m}_{\lambda}(z)\right] \label{tyty1}    \\  
&=& (\lambda -\m+1) (\lambda +\m) \P^{-\m}_{\lambda}(z) .  \label{popopol}  
\end{eqnarray}

It takes a little work to verify that the above formulae are nothing but a rewriting of known relations among the Legendre functions. To prove Eq. (\ref{tyty1}) one first removes the derivative  ${\P'}= \frac {d\P}{dz} $ by using  Eq. 3.8.19 from Bateman's book  \cite{HTF1} and get
   \begin{eqnarray}
&&  i ( \lambda -\m+1)\sinh t \,  \, {\P^{1-\m}_{\lambda}}(z)+ 
 { (\lambda-\m+1) }  {\P^{1-\m}_{\lambda-1}}(z)
  \cr && -  (\lambda -\m+1) (\lambda +\m) \left[ - i \sinh t \,(\lambda\,+\m+1)  {\P^{-1-\m}_{\lambda}}(z)+ { (\lambda-\m-1) }  {\P^{-1-\m}_{\lambda-1}}(z) \right] =0.\label{pol11}
   \end{eqnarray}
Eqs. 3.8.11 and 3.8.15 from  allow to show that (\ref{pol11}) is equivalent to 
    \begin{eqnarray}
  {\P^{1-\m}_{\lambda}} (z)  - 2\m\,  i  \tanh t  \,  \P^{-\m}_{\lambda} (z) +  (\lambda -\m)  \,(\lambda\,+\m+1)  {\P^{-1-\m}_{\lambda}}(z) =0 \label{nm}
   \end{eqnarray}  
 which in turn coincides with  Eq. 3.8.11 from \cite{HTF1}.  
To prove the second equality (\ref{popopol}) one invokes Eqs. 3.8.17 and  3.8.19  \cite{HTF1}.

Now we are ready to show  the $\slr$-invariance of the kernel (\ref{tpabh0}). Let us insert Eqs. (\ref{tyty1})  and (\ref{popopol}) in Eq. (\ref{condf}) and divide by  $\gamma_{\m-1}$; we get the following equivalent expression
\begin{eqnarray}
&&[ e^{i\theta'}\  \P^{1-\m}_{\lambda}(-z' )   - e^{-i\theta'} (l-\lambda ) (\lambda +l+1) \  \P^{-1-\m}_{\lambda}(-z' )]    +\cr
  &&+ 2 \left[ i \sin \theta' \cosh t' \,  {{\P'}^{-\m}_{\lambda}}(-z')+ i \m  \tanh t'\, \cos \theta'\,  \P^{-\m}_{\lambda}(-z') \right]=0. 
\end{eqnarray}
Here the variable $z$ has disappeared and the condition (\ref{condf}) is now tractable. By singling out the coefficients of $\cos \theta'$ and $\sin \theta'$  we are led to examine the validity of the following identities:
\begin{eqnarray}
&& \P^{1-\m}_{\lambda}(-z' )   - (l-\lambda ) (\lambda +l+1) \  \P^{-1-\m}_{\lambda}(-z' )   + 2  i \m  \tanh t'\,  \P^{-\m}_{\lambda}(-z') =0, \label{firsta}
\\
&&  \P^{1-\m}_{\lambda}(-z' )   +   (l-\lambda ) (\lambda +l+1) \  \P^{-1-\m}_{\lambda}(-z' )    + 2 \cosh t' \,  {{\P'}^{-\m}_{\lambda}}(-z') =0. \label{firstb1}
\end{eqnarray}
Eq. (\ref{firsta}) it is once more a known relationship among 
Legendre functions on the cut, namely Eq. 3.8.11 of \cite{HTF1}.
As regards the second identity, it can be proven by observing that   the difference of the above two equations 
coincides with the relation given in Eq. (\ref{tyty1}). The  $\slr$-invariance of the kernel (\ref{tpabh0}) is proven. 

\vskip 10 pt

An immediate  corollary that follows from Eqs. (\ref{firsta}) and (\ref{firstb1}) is that  the kernels obtained by taking the even and the odd parts of $\twofcn_0 (\tilde x,\tilde x') $, namely 
\begin{eqnarray}
&& \twofcn_{0,even} (\tilde x,\tilde x') = \frac 1 {4\pi}\sum_{l \in {\Bbb Z}}   \gamma_l
\left[ \P^{-l}_{\lambda}(z)
\P^{-l}_{\lambda}(-z') \right]  \exp(il(\theta-\theta'))\label{tpabh0s}
\end{eqnarray}
and
\begin{eqnarray}
&& \twofcn_{0,odd} (\tilde x,\tilde x') = \frac 1 {4\pi}\sum_{\frac 12+l  \in {\Bbb Z}}   \gamma_l
\left[ \P^{-l}_{\lambda}(z)
\P^{-l}_{\lambda}(-z') \right]  \exp(il(\theta-\theta')) \label{tpabh0a}
\end{eqnarray}
are separately invariant. The even part coincides with the so-called Bunch-Davies vacuum \cite{Bunch,Chernikov,hawking,bm,bgm,bem} .

\vskip 10 pt

The second corollary is the invariance of the commutator.
Let us consider indeed the map 
$
\tau_1$ 
given in Eq. (\ref{m.35.a}).
Since it commutes with the action of $\slr$  on the covering manifold $\widetilde {dS_{2}}$ we immediately get that also the kernel
\begin{eqnarray}
&& \twofcn_1 (\tilde x, \tilde x') = \sum_{2l \in {\Bbb Z}}   \gamma_l
\left[ \P^{-l}_{\lambda}(i\sinh t)
\P^{-l}_{\lambda}(-i\sh t') \right]  \exp(-il(\theta-\theta')) \label{tpabh}
\end{eqnarray}
is $\slr$ invariant. The invariance of the commutator follows.

\section{Invariance under $\slr$ and other properties of general
two-point functions}
\label{proof}

Once given the commutator, the crucial step to get a physical model is to represent 
the field $\phi$ as an operator-valued distribution in a
Hilbert space $\cal H$.  This can be done 
by finding a positive-semidefinite
bivariate distribution ${\twofcn}(\tilde x,\tilde y)$  solving the KG equation and the functional equation \cite{ms,ms2,sw}
\begin{equation}
{C}(\tilde x,\tilde x') = {\twofcn}(\tilde x,\tilde x')- {\twofcn}(\tilde x',\tilde x). \label{CR}
\end{equation}
Actually, $C$ and $\twofcn$ are not functions but distributions so the above equation must be understood in the sense of distributions.  
Given  a solution  $\twofcn$ of Eq. (\ref{CR}) an analogue of Wightman's reconstruction theorem \cite{sw} (in the simplest case of generalised free fields)  provides the
Fock space of the theory and a  representation of the field as a local operator-valued distribution.

There are of course infinitely many inequivalent solution of Eq. (\ref{CR}). 
Here we will characterize the most general  $\slr$--invariant solution. 
To this aim let us first  consider a general two-point function, i.e. a distribution ${\twofcn}$ on
$\wt{dS_2}\times \wt{dS_2}$. There are several conditions that we may (or may not)
want to impose on such a function.
\begin{description}

\item[(C1)]  Local commutativity (locality):
Let
\beq
\RR = \{(\wtx,\ \wtx') \in \wt{dS_2}\times \wt{dS_2}\ :\ 
\makebox{$\wtx$ and $\wtx'$ are spacelike separated} \} .
\label{p.9}\endq
$\twofcn(\wtx, \wtx')$ has the property of local commutativity
(or locality) if
\beq
\twofcn(\wtx, \wtx') - \twofcn(\wtx', \wtx) = 0 \ \ \ \
\forall (\wtx,\wtx') \in \RR\ .
\label{p.10}\endq

\item[(C2)]  Symmetry or anti-symmetry:
\beq
\twofcn(\wtx, \wtx') = \pm \twofcn(\wtx', \wtx)\ \ \ \ \forall (\wtx,\wtx') \in \wt{dS_2}\times \wt{dS_2} .
\label{p.11}\endq

\item[(C3)]  Invariance under the group $\slr$ :
\beq
\twofcn(\tilde g \wtx,\tilde g \wtx') = \twofcn(\wtx,\wtx')  \ \ \ \ \forall (\wtx,\wtx') \in \wt{dS_2}\times \wt{dS_2} .
\ \ \forall \tilde g \in \slr.
\label{p.15}\endq

\item[(C4)]  Hermiticity:
\beq
\twofcn(\wtx,\wtx') = \ovl{\twofcn(\wtx',\wtx)}   \ \ \ \
\forall (\wtx,\wtx') \in \wt{dS_2}\times \wt{dS_2} .
\label{p.20}\endq

\item[(C5)]  Positive definiteness: 
\beq
\int \int \twofcn(\wtx,\wtx') \bar{f} (\wtx) f(\wtx') d\wtx d\wtx' \geq 0 \ \ \ \  \forall f \in {\cal C}_0^\infty (\wt{dS_2}) . \label{posdef}
\endq
\item[(C6)]  Klein-Gordon equation in $\wtx$ and $\wtx'$ with a ``mass'' $\lambda$.
 \item[(C7)]  Canonical Commutation Relations (\ref{CR}). 
 \item[(C8)]  Analyticity.
By this we mean that there is an open tuboid\footnote{See a general
discussion of tuboids in \cite{bem}.}
$\UU_+$ in a complexified
version of $\wt{dS_2}\times \wt{dS_2}$ (see Sect. \ref{coset})
such that, in a neighborhood of
any real point $(x,\ x') \in \wt{dS_2}\times \wt{dS_2}$ we have,
in the sense of distributions,
\beq
F(x,\ x') = \lim_{(w,\ w')\in \UU_+\,,\ (w,\ w')
\rightarrow (x,\ x')} F_+(w,\ ,w')\ ,
\label{p.50}\endq
where $F_+$ is holomorphic with locally polynomial behavior in $\UU_+$.
$F_-(w,\ w')\bydef F_+(w',\ w)$ is analytic in
\beq
\UU_- = \{(w,\ w')\ :\ (w',\ w)\in \UU_+\}
\label{p.51}\endq
and we suppose
\beq
\UU_- = \ovl{\UU_+}\ .
\label{p.52}\endq

\item[(C9)] Local analyticity:

By this we mean that there is a complex open
connected neighborhood $\NN$ of $\RR$ such that, in $\RR$, both
$F(\wtx,\ \wtx')$ and $F(\wtx',\ \wtx)$ are restrictions of the
same function holomorphic in $\NN$. If a two-point function $F$
has the two properties of locality and analyticity as defined above,
then it also has the property of local analyticity by the edge-of-the-wedge
theorem.

\end{description}

If $\twofcn$ is any two-point function, it can be written as $\twofcn = \twofcn_r + i\twofcn_i$, where
\beq
 \twofcn_r(\wtx,\wtx') = \haf \twofcn(\wtx,\wtx') + \haf \ovl{\twofcn(\wtx',\wtx)} ,
\ \ \ \twofcn_i(\wtx,\wtx') = {1\over 2i} \twofcn(\wtx,\wtx') -
{1\over 2i}\ovl{\twofcn(\wtx',\wtx)}\ .
\label{p.21}\endq     
$\twofcn_r$ and $\twofcn_i$ are hermitic, and if $\twofcn$ satisfies any one of the
conditions (C2), (C3), or (C6), so do $\twofcn_r$ and $\twofcn_i$.
If ${\twofcn}$ is any two-point function, it can be written as $\twofcn = \twofcn_{ev en}+ \twofcn_{odd}$ where 
\beq
\twofcn_{even}(\wtx,\wtx') = \haf \twofcn(\wtx,\wtx') + \haf \twofcn(\wtx, \tau \wtx'),\ \ \ 
\twofcn_{odd}(\wtx,\wtx') = \haf \twofcn(\wtx,  
\wtx') - \haf \twofcn(\wtx, \tau \wtx')\ .
\label{p.22}\endq

Those two-point functions which
are rotation-invariant in our fixed frame, i.e. such that 
\beq
\twofcn((t, \theta),(t', \theta')) = \twofcn((t, \theta+a),(t', \theta'+a))
\ \ \ \forall a\in \bR\ 
\label{m1.60}\endq
(recall that $\tau$ is such a rotation for $a= 2\pi$)
 can be expanded in a Fourier series as follows:
\beq
\twofcn(\wtx,\wtx') = \sum_{l\in \LL} u_l(z,z') e^{il(\theta-\theta')}\ .
\label{f.10}\endq
As before  $z = i\sh t $ and $z' = i\sh t'$.
The set $\LL$ can be ${\Bbb Z}$, $\haf{\Bbb Z}$, or $\haf +{\Bbb Z}$.
If $\LL = \half \bZ$ then $\twofcn_{even}$ (resp. $\twofcn_{odd}$), as
defined in (\ref{p.22}), is the sum over $\bZ$ (resp. $\half+ \bZ$). 

The following is the main result of the present paper:

\begin{theorem}
\label{fourier} 
The most general  $\slr$--invariant  hermitic two-point function  satisfying the Klein-Gordon equation in each variable for a positive squared mass 
$m^2_\lambda = - \lambda(\lambda+1)>0$ (i.e. for the principal and the complementary values) is characterized  by four independent real constants  $A_0, B_0, A_{\frac 12}, B_{\frac 12}$ and two independent complex constants $C_0, C_{\frac 12}$ by the following Fourier expansion

\begin{align}
\twofcn(\wtx,\wtx') &= \sum_{2 l\in {\Bbb Z}} \, \gamma_l \, [ A_l \P^{- l }_{\lambda}(z)\P^{- l }_{\lambda}(-z')
+B_l \P^{- l }_{\lambda}(-z)\P^{- l }_{\lambda}(z')\cr
&+ e^{i\pi l}C_l \P^{- l }_{\lambda}(z)\P^{- l }_{\lambda}(z')
+ e^{-i\pi l}C_l^* \P^{- l }_{\lambda}(-z)\P^{- l }_{\lambda}(-z') ] e^{il(\theta-\theta')}. 
\label{wig}
\end{align}
where 
\begin{eqnarray}
&&  A_l = A_0,\ \ B_l = B_0, \ \ C_l = C_0  \ \ for \ \  l \in {\Bbb Z}  \label{inve0} \cr
&& A_l = A_{\frac 12 },\ \ B_l = B_{\frac 12}, \ \ C_l = C_{\frac 12} \ \  for   \ \ l \in \frac 12 + {\Bbb Z}. \nonumber
\end{eqnarray}
\end{theorem}

\vskip20 pt
 {\em Here we neither impose the locality property nor the positive definiteness}. We have put  $\gamma_l
= \frac 12 {\Gamma(l-\lambda) \Gamma(1 +\lambda + l ) } $ in evidence for convenience, by taking inspiration from the previous section.

Let us therefore consider the kernel given in Eq. (\ref{wig}). For the chosen mass parameters $\lambda$  and $z \in i\bR$ it happens that 
$\ovl{\P_\lambda^{-l}(z)} = \P_\lambda^{-l}(-z)$. With this restriction,
$W$ is hermitic iff $A_l = A_l^*$, $B_l = B_l^*$.

The two-point function (\ref{wig}) is  $\slr$-invariant if and only if  condition (\ref{conddd}) holds.
This amounts to 
\begin{eqnarray}
\begin{array}{r}
\delta_A+ \delta_B+\delta_C+\delta_{C^*}=\sum_{2l\in \bZ}  i \gamma_l  \sin \theta \cosh t \left[ (A_l  {\P^{-l}_{\lambda}}(-z')+
C_l e^{i\pi l } {\P^{-l}_{\lambda}}(z'))
 {{\P^\prime}^{-l}_{\lambda}}(z)   \right.\cr - \left. (B_l  {\P^{-l}_{\lambda}}(z')+ 
C^*_l e^{-i\pi l } {\P^{-l}_{\lambda}}(-z')) {{\P^\prime}^{-l}_{\lambda}}(-z)]\right]e^{il(\theta-\theta')} 
\cr
+\sum_{2l\in \bZ}  i l \gamma_l      \tanh t\, \cos \theta\,\left[  (A_l  {\P^{-l}_{\lambda}}(-z')+
C_l e^{i\pi l } {\P^{-l}_{\lambda}}(z')) {\P^{-l}_{\lambda}}(z)  \right.\cr\left. +  (B_l  {\P^{-l}_{\lambda}}(z')+ 
C^*_l e^{-i\pi l } {\P^{-l}_{\lambda}}(-z')) {\P^{-l}_{\lambda}}(-z)\right]e^{il(\theta-\theta')} 
\cr
+\sum_{2l\in \bZ}  i \gamma_l  \sin \theta' \cosh t'  \left[ (B_l  {\P^{-l}_{\lambda}}(-z)+
C_l e^{i\pi l } {\P^{-l}_{\lambda}}(z))
 {{\P^\prime}^{-l}_{\lambda}}(z')   \right.\cr\left.  - (A_l  {\P^{-l}_{\lambda}}(z)+ 
C^*_l e^{-i\pi l } {\P^{-l}_{\lambda}}(-z)) {{\P^\prime}^{-l}_{\lambda}}(-z') \right]e^{il(\theta-\theta')}
 \cr
-  \sum_{2l\in \bZ}  i l  \gamma_l  \tanh t'\, \cos \theta'\, \left[  [(B_l  {\P^{-l}_{\lambda}}(-z)+
C_l e^{i\pi l } {\P^{-l}_{\lambda}}(z)) {\P^{-l}_{\lambda}}(z')  \right.\cr\left. + (A_l  {\P^{-l}_{\lambda}}(z)+ 
C^*_l e^{-i\pi l } {\P^{-l}_{\lambda}}(-z)) {\P^{-l}_{\lambda}}(-z')]\right]e^{il(\theta-\theta')} = 0 
\end{array} \label{000}
\end{eqnarray}
where $\delta_A$ includes all  the terms containing $A$ and so on. Singling out the Fourier coefficient of $\exp i l \theta$ we get 
\begin{eqnarray}
\delta_A(l)
&=&
\frac 12 e^{-il\theta'} \gamma_{\m-1}e^{i\theta'}\ A_{\m-1}   \P^{1-\m}_{\lambda}(-z' )   \left[  \cosh t \,  {{\P'}^{1-\m}_{\lambda}}(z)+ i (\m-1)  \tanh t\, \,  \P^{1-\m}_{\lambda}(z) \right]
+\cr 
 &+& \frac 12 e^{-il\theta'} \gamma_{\m+1} e^{-i\theta'}\ A_{\m+1}   \P^{-1-\m}_{\lambda}(-z' )   \left[  -\cosh t \,  {{\P'}^{-1-\m}_{\lambda}}(z)+i (\m+1)  \tanh t\, \,  \P^{-1-\m}_{\lambda}(z) \right]
+\cr
 &-& e^{-il\theta'} \gamma_l   A_{\m} \P^{-\m}_{\lambda}(z )  \left[ i \sin \theta' \cosh t' \,  {{\P'}^{-\m}_{\lambda}}(-z') + i \m \cos \theta'  \tanh t'\,  \P^{-\m}_{\lambda}(-z') \right] 
  \label{condda}
\end{eqnarray}
By taking into account  the crucial identities    (\ref{tyty1}) and (\ref{popopol})  and also Eq. (\ref{gammaal})  this expression takes the following  simpler form:
\begin{eqnarray}
\delta_A(l)
&=& - \frac 12 e^{-il\theta'} \gamma_{\m}  e^{i\theta'}\ A_{\m-1}   \P^{1-\m}_{\lambda}(-z' )   
 \P^{-\m}_{\lambda}(z) 
 \cr 
 &+& \frac 12 e^{-il\theta'} \gamma_{\m}  (l-\lambda ) (\lambda +l+1) e^{-i\theta'}\ A_{\m+1}   \P^{-1-\m}_{\lambda}(-z' )   {{\P}^{-\m}_{\lambda}}(z) 
+\cr
 &+& e^{-il\theta'} \gamma_l   A_{\m} \P^{-\m}_{\lambda}(z )  \left[ i \sin \theta' ( i \m  \tanh t'\, \,  \P^{-\m}_{\lambda}(-z')   +  \P^{1-\m}_{\lambda}(-z'))
 - i \m \cos\theta' \tanh t'\,  \P^{-\m}_{\lambda}(-z') \right]. 
  \label{conddabis}\cr &&
\end{eqnarray}
By using the operator $\tau_1$ and $\tau_2$ we also immediately get that 
\begin{eqnarray}
\delta_B(l)
&=&  \frac 12 e^{-il\theta'} \gamma_{\m}  e^{i\theta'}\ B_{\m-1}   \P^{1-\m}_{\lambda}(z' )   
 \P^{-\m}_{\lambda}(-z) 
 \cr 
 &-& \frac 12 e^{-il\theta'} \gamma_{\m}  (l-\lambda ) (\lambda +l+1) e^{-i\theta'}\ B_{\m+1}   \P^{-1-\m}_{\lambda}(z' )   {{\P}^{-\m}_{\lambda}}(-z) 
+\cr
 &-& e^{-il\theta'} \gamma_l   B_{\m} \P^{-\m}_{\lambda}(-z )  \left[ i \sin \theta' ( i \m  \tanh t'\, \,  \P^{-\m}_{\lambda}(z')   +  \P^{1-\m}_{\lambda}(z'))
 - i \m \cos\theta' \tanh t'\,  \P^{-\m}_{\lambda}(z') \right]  ,
  \label{conddabiso} \cr && \cr
\delta_C(l)
&=&  \frac 12 e^{-il\theta'}  e^{i\pi \m } \gamma_{\m}  e^{i\theta'}    C_{\m-1}   \P^{1-\m}_{\lambda}(z' )   
 \P^{-\m}_{\lambda}(z) 
 \cr 
&-& \frac 12 e^{-il\theta'}  e^{i\pi \m }\gamma_{\m}  (l-\lambda ) (\lambda +l+1) e^{-i\theta'} C_{\m+1}   \P^{-1-\m}_{\lambda}(z' )   {{\P}^{-\m}_{\lambda}}(z) 
+\cr
 &-& e^{-il\theta'}e^{i\pi \m }  \gamma_l  C_{\m} \P^{-\m}_{\lambda}(z )  \left[ i \sin \theta' ( i \m  \tanh t'\, \,  \P^{-\m}_{\lambda}(z')   +  \P^{1-\m}_{\lambda}(z'))
 - i \m \cos\theta' \tanh t'\,  \P^{-\m}_{\lambda}(-z') \right], 
  \label{conddabisok} \cr && \cr
\delta_{C^*}(l) &=& - \frac 12 e^{-il\theta'} e^{-i\pi \m }  \gamma_{\m}  e^{i\theta'}\ C^*_{\m-1}   \P^{1-\m}_{\lambda}(-z' )   
 \P^{-\m}_{\lambda}(-z) 
 \cr 
 &+& \frac 12 e^{-il\theta'} e^{-i\pi \m } \gamma_{\m}  (l-\lambda ) (\lambda +l+1) e^{-i\theta'}\ C^*_{\m+1}   \P^{-1-\m}_{\lambda}(-z' )   {{\P}^{-\m}_{\lambda}}(-z) 
+\cr
 &+& e^{-il\theta'} e^{-i\pi \m }  \gamma_l   C^*_{\m} \P^{-\m}_{\lambda}(-z )  \left[ i \sin \theta' ( i \m  \tanh t'\, \,  \P^{-\m}_{\lambda}(-z')   +  \P^{1-\m}_{\lambda}(-z'))
 - i \m \cos\theta' \tanh t'\,  \P^{-\m}_{\lambda}(-z') \right]. 
  \label{conddabismm}\cr && \nonumber
\end{eqnarray}
Let us begin by discussing the simpler case where $B=C=0$ which reduces to $\delta_A(l)=0$.
Singling out as before the coefficients of  $\sin\theta'$ and $\cos \theta'$ we get
\begin{eqnarray}
\begin{array}{l}
\left(\left(A_l - \frac {A_{\m-1} }2 \right) \P^{1-\m}_{\lambda}(-z' )   -  (l-\lambda )  (\lambda +l+1) \frac {A_{\m+1}}2     \P^{-1-\m}_{\lambda}(-z' ) 
  + i   \m A_{\m}   \tanh t'  \,  \P^{-\m}_{\lambda}(-z')\right)  \P^{-\m}_{\lambda}(z)   =0    
\\ \\
\left(- \frac{ A_{\m-1}}2    \P^{1-\m}_{\lambda}(-z' )   +  (l-\lambda)  (\lambda +l+1)  \frac {A_{\m+1}}2     \P^{-1-\m}_{\lambda}(-z' )   -  i \m A_l  \tanh t'   \P^{-\m}_{\lambda}(-z')\right)  \P^{-\m}_{\lambda}(z)  = 0.
\end{array}\nonumber
\end{eqnarray}
Taking their sum we get 
\begin{eqnarray}
\left(A_l - A_{\m-1} \right) \P^{1-\m}_{\lambda}(-z' ) \P^{-\m}_{\lambda}(z)   =0 .
\end{eqnarray}
This shows that $A_l = A_0$ for  $l\in {\Bbb Z}$  and  $A_l = A_{\frac 12 }$ for $l \in \frac 12 + {\Bbb Z}$.
The above two equations now reduce to 
\begin{eqnarray}
&&  \P^{1-\m}_{\lambda}(-z' )   - (l-\lambda)  (\lambda +l+1)     \P^{-1-\m}_{\lambda}(-z' )   +2  i \m  \tanh t'   \P^{-\m}_{\lambda}(-z')= 0 
\end{eqnarray}
and this is a known relation between contiguous Legendre functions (Bateman Eq. 3.8.11).
In the general case, proceeding in the same way, we get that condition (\ref{000}) implies 
\begin{eqnarray}
&& \left(A_l - A_{\m-1} \right) \P^{1-\m}_{\lambda}(-z' ) \P^{-\m}_{\lambda}(z)   -   \left(B_l - B_{\m-1} \right) \P^{1-\m}_{\lambda}(z' ) \P^{-\m}_{\lambda}(-z)  + \cr 
&-& e^{i\pi l}\left(C_l - C_{\m-1} \right) \P^{1-\m}_{\lambda}(z' ) \P^{-\m}_{\lambda}(z)   +  e^{-i\pi l} \left(C^*_l - C^*_{\m-1} \right) \P^{1-\m}_{\lambda}(-z' ) \P^{-\m}_{\lambda}(-z)  = 0 . 
\end{eqnarray}
Therefore also in the general case the  possible values for $A_l$,  $B_l$ and $C_l$ are
characterized by only 8 arbitrary real constants:
\begin{eqnarray}
&&  A_l = A_0,\ \ B_l = B_0, \ \ C_l = C_0  \ \ for \ \  l \in {\Bbb Z}  \label{inve} \\
&& A_l = A_{\frac 12 },\ \ B_l = B_{\frac 12}, \ \ C_l = C_{\frac 12} \ \  for   \ \ l \in \frac 12 + {\Bbb Z}. \label{invo}
\end{eqnarray}
The verification that these conditions indeed guarantee that $\delta_A(l) + \delta_B(l) + \delta_C(l) + \delta_{C^*}(l)=0$ proceeds as before.

\subsection{Canonicity}
If we impose that an invariant two-point function as in the previous theorem satisfies also the
canonical commutation relation (\ref{CR}),
a  simple calculation shows that 
\begin{equation}
A_0-B_0 = {1\over 2k\pi}, \ \ A_\haf = {1\over 4k\pi} - e^{i\pi l} C_\haf,
\ \, B_\haf = -{1\over 4k\pi} - e^{i\pi l} C_\haf, \ \ C_\haf = -C_\haf^*
\end{equation}
while $C_0$ is unrestricted ($A_\haf$, $B_\haf$ and $C_\haf$ are present
only if $\LL = \haf \bZ$).

\subsection{Positivity}
\label{positivity}
Let us again consider a hermitic 2-point function $W$ as in Eq. (\ref{f.10}).
A necessary and sufficient condition for
(C5) (Eq. (\ref{posdef})) to hold is that, for every $l$
and every test function $f$ on $\bR$,
\beq
\int_{\bR\times \bR}\ovl{f(t)}u_l(z,\ z')f(t')\,dt\,dt' \ge 0\ ,
\label{u.33}\endq
where $z = i\sh t$, $z' = i\sh t'$.
Suppose that $W$ is of the form (\ref{wig}) with $A_l=A_l^*$, $B_l=B_l^*$.
If $f$ is a test-function on $\bR$, let
\beq
f_1 = \int_\bR f(t)\P_\lambda^{-l}(- i\sh t )\, dt,\ \ \ 
f_2 = \int_\bR f(t)\P_\lambda^{-l}(i\sh t )\, dt.
\label{u.50}\endq
Then
\beq
\int_{\bR\times\bR} \ovl{f(t)}u_l(z,\ z') f(t')\,dt\,dt' = \gamma_l
\left ( \begin{array}{cc}
\ovl{f_1} & \ovl{f_2} \end{array} \right )
\left ( \begin{array}{cc}
A_l & e^{i\pi l}C_l\\ e^{-i\pi l}C^*_l & B_l \end{array} \right )
\left ( \begin{array}{c}
f_1\\ f_2 \end{array} \right )\ .
\label{u.52}\endq
Therefore $\twofcn$ is of positive type if and only if
\beq
\gamma_l A_l \ge 0,\ \ \gamma_l B_l \ge 0,\ \
\gamma_l^2 (A_l B_l - C_l C^*_l) \ge 0\ \ \ \forall l \in \LL\ .
\label{u.53}\endq
Let us now suppose that $\twofcn$ is invariant, i.e. the conditions
(\ref{inve}) and (\ref{invo}) are satisfied. If $l\in \bZ$,
with our choices of $\lambda$, $\gamma_l > 0$ for all $l\in \bZ$,
hence $\twofcn_{even}$ is of positive type iff
\beq
A_0 > 0,\ \ B_0 >0,\ \ A_0B_0 -C_0 C^*_0 > 0.
\label{u.53.1}\endq
If $l \in \haf +\bZ$ and $\lambda = -\haf + i\rho$, $\rho \not= 0$,
then $\gamma_l$ is always $> 0$ and $\twofcn_{odd}$ is of positive type iff
\beq
A_\haf > 0,\ \ B_\haf >0,\ \ A_\haf B_\haf -C_\haf C^*_\haf > 0.
\label{u.53.2}\endq
If $l \in \haf +\bZ$ and $-1 < \lambda < 0$,
then $\gamma_l$ has the sign of $l$ and
$\twofcn_{odd}$ is {\em never of positive type}. 

The  consquence of the last statement is the {\em disappearance of the complementary series on the double covering of the de Sitter manifold}. In other words there exists no local and $\slr$--covariant scalar free field on $\wt{dS_2}$ with mass 
\begin{equation}
0<m^2<\frac 14.
\end{equation}
We will clarify further this point by studying the vacuum representations in the following section.

\section{Study of local "vacuum" states invariant under $\slr$}

One particular instance of Eq. (\ref{wig}) is the two-point function constructed in terms of the system of modes $\phi_l$ (\ref{modesab01})
as their "vacuum" expectation value: 
\begin{eqnarray}
&&\twofcn (x,x') = \sum_l \phi_l(x) \phi_l^*(x') = \cr &= &
\sum_l  [a_l\P^{- l }_{\lambda}(i\sh t) +b_l  \P^{- l }_{\lambda}(-i\sh t)]
[a^*_l\P^{- l }_{\lambda}(-i\sh t') +b^*_l  \P^{- l }_{\lambda}(i\sh t')]e^{{il \theta-il \theta'}}  \label{tp0}  
\end{eqnarray}
For any possible choice of $a_l$ and $b_l$ the "vacuum states" given by Eq.  (\ref{tp0}) are   "pure states" i.e. they provide through the GNS construction 
irreducible representations of the field algebra.

Let us now single out among them those states who are $\slr$--invariant.
This is an easy corollary of the theorem of the previous section. Eqs. (\ref{inve}) and (\ref{invo})  imply the following relations:
\begin{eqnarray}    |a_l|^2 }= c_1(\epsilon) \gamma_l,   \ \  {|b_l|^2 = c_2(\epsilon)  \gamma_l , \ \    a_lb_l^* = c_3(\epsilon)  \gamma_l e^{i l\pi}\label{ccond} \end{eqnarray}
where $\epsilon = 0$ for $l\in{\Bbb Z}$ and  $\epsilon = 1$ for $l\in \frac 12+{\Bbb Z}$ (i.e. there are six independent constants).
The above equations, together with the normalization condition (\ref{kkgg12}), can be solved as follows:
\begin{eqnarray}
&& a_{ l} = \sqrt{ \frac {\gamma_l} {2\pi k}} \cosh \alpha_\epsilon, \ \ \ \  b_{ l} = \sqrt{ \frac {\gamma_l } {2\pi k}} \sinh \alpha_\epsilon e^{i\phi_\epsilon-il\pi}, \ \ \ \ \epsilon=0, 1. \label{alpha}
\end{eqnarray}
Here we took $a_l$ real without loss of generality.
\subsection{Canonicity: pure de Sitter}
Let us examine whether the above equations are compatible with the requirements imposed by canonicity. 
In the pure de Sitter case (as opposed to its covering)  $l$ is integer and the CCR's amount to the condition (\ref{kgpds})
which imposes no further restriction 
and any choice of $\alpha_0$ and $\phi_0$ gives rise to a de Sitter invariant 
state which has the right commutator (relatively to the de Sitter manifold). 
These states are well-known: they are the so-called alpha vacua \cite{spindel,allen,mottola,tach}. 

Among them, there is a particularly important state corresponding to the choice $\alpha_0= 0$: this is the so-called Bunch-Davies vacuum \cite{Bunch,Chernikov,hawking,bm,bgm,tach} 
\begin{eqnarray}
\twofcn_{BD}(x,x')=  \twofcn^{(0)}_{\alpha_0=0}(x,x')= \sum_{l \in {\Bbb Z}}\frac { \gamma_l }{2\pi} \P^{- l }_{\lambda}(i\sh t) 
\P^{- l }_{\lambda}(-i\sh t') e^{{il \theta-il \theta'}} = \cr =
{\Gamma(-\lambda)\Gamma(\lambda+1)\over 4\pi}
\,P_{\lambda }(\zeta),
 \label{tbd}  
 \end{eqnarray}
where $P_\lambda(\zeta) $ is the associated Legendre function of the first kind \cite{HTF1} and the  de Sitter invariant variable $\zeta$ is the scalar product $\zeta = x(t-i\epsilon,\theta) \cdot x'(t'+i\epsilon,\theta)$ in the ambient space sense.
Actually, $\twofcn_{BD}(x,x')$ admits an extension to the complex de Sitter manifold and satisfies there  the {\em maximal analyticity property} \cite{bm,bgm,bem}:
it is holomorphic for all $\zeta\in \bC\setminus (-\infty,\ -1]$ i.e. everywhere except on the locality cut. 
This crucial property singles the Bunch-Davies vacuum out of all the other invariant vacua and has a very well known thermal interpretation \cite{hawking,bm,bgm,bem}: the restriction of the Bunch-Davies state a wedge-like region is a thermal state at temperature $T=1/2\pi$. A similar property is expected in interacting theories based on an analogue of the Bisognano-Wichmann theorem \cite{bem}. We will come back on this point later.

\subsection{Covering}
In the antiperiodic case the CCR's 
\begin{eqnarray}
&&\makebox{for $l\in\frac 12 +{\Bbb Z}$} \ \  \left\{\begin{array}{l}
a_l  a_{-l}  -b_l  b_{-l} = c_l\sin (\pi  \lambda ) \\  a_l  b_{-l}  -b_l  a_{-l}  =c_l \sin(\pi l)\end{array} \right. \end{eqnarray}
imply the following  relation between the constants $\alpha$ and $ \phi$ and the mass parameter $\lambda$ of the field:  
\beq
e^{i \phi } \sin (\pi  l) (-i \sinh (2 \alpha ) \sin (\pi  \lambda )-i \cosh (2 \alpha )
   \sin \phi +\cos \phi )=0. \label{condoo}
\end{equation}
 For $\lambda = -1/2+ i\nu$ there is only one possible solution  given by   
\begin{equation}
\coth 2 \alpha = \cosh \pi \nu, \ \ \ \phi = \frac \pi 2.
\end{equation}
We denote the corresponding two-point function $\twofcn^{(\frac 12) }_{i\nu}(x,x')$. 
Note that the value $\alpha =0$,  that would correspond to the above-mentioned maximal analyticity property, is excluded: it would be attained only for an infinite value of the mass. 
On the other hand  Equation (\ref{condoo}) has no solution at all when  $\lambda$ is real: there is no invariant vacuum of the complementary series.

In conclusion, for $\lambda = -\frac 12 + i \nu$  the most general  invariant vacuum state is the superposition of an arbitrary alpha vacuuum (the even part) plus a fixed odd part  $\twofcn^{(\frac 12) }_{\nu}(x,x')$ as follows 
\begin{equation}
\twofcn(x,x') = \twofcn^{(0)}_{\alpha_0, \phi_0}(x,x') +  \twofcn^{(\frac 12) }_{i \nu}(x,x').
\end{equation}
For $\lambda = -\frac 12 + \nu$ there is no $\slr$ local invariant vacuum state i.e. there is no field of a would-be complementary series.

\vskip 20 pt

\newpage

\section{Convergence and (lack of) analyticity}
\label{conver}
We again consider a series of the form (\ref{wig}). In the preceding
sections such a series was regarded as the Fourier expansion of
some two-point function (or rather distribution). Here
we suppose the series given and ask about its convergence. No generality
is lost by the restriction to a hermitic series. However we will consider
only a particular example from which the general case can be understood.
Let
\begin{align}
F_0(x,x') = \sum_{l\in \LL}c_l(z,z')e^{il(\theta-\theta')}\ ,\ \ \ \ \ 
c_l(z,z') = \gamma_l \P_\lambda^{-l}(z)\P_\lambda^{-l}(-z')\ ,
\label{cv.10}\end{align}
where $z= i\sh t$ and $z'= i\sh t'$.
The dependence of $c_l$ on $\lambda$ has been omitted for simplicity,
and $\LL = \haf\bZ$.
This series is the simplest example of the $\slr$ invariant
series discussed in Theorem (\ref{fourier}) .
We also set
\begin{align}
&z = ix^0 = i\sh t =  i \tg s,\ \  
z' = ix'^0 = i\sh t' = i  \tg s' \cr
&  u = s+\theta, \ \  v = s-\theta,\ \ u' = s'+\theta',\ \ v' = s'-\theta'.\ 
\label{m.25}\end{align}
Note that the variables $u$, $v$, $u'$ $v'$ defined here and used in this
section are not those used in Sect.~\ref{ds2coset}.
$c_l(z, z')$ is holomorphic in $z$ (resp. $z'$) in the
cut-plane $\Delta_2$ (see (\ref{t.22})). Values such that
$\Re z > 0,\ \Re z' <0$ correspond to $x\in \TT_-\,,\ x'\in \TT_+$
while $\Re z < 0,\ \Re z' >0$ correspond to $x\in \TT_+\,,\ x'\in \TT_-$.
The convergence of the series can be studied separately for
$l \in \bZ$ and $l\in \haf+ \bZ$.

\subsection{Case of integer $l$}
\label{convint}
We wish to investigate the convergence of the series (\ref{cv.10}) in the
case when $l \in \bZ$, $\Re z >0$ and $\Re z' <0$. 
One may check that {\it if $l\in \bZ$, then $c_l(z,\ z') = c_{-l}(z,\ z')$}.
The proof is based on  the following relation 
\cite[3.4 (17) p. 144]{HTF1}
\beq
\P_\lambda^l(z) = {\cos(l\pi)\Gamma(\lambda+l+1)\over
\Gamma(\lambda-l+1)}\P_\lambda^{-l}(z)\ ,
\label{m.35}\endq
Therefore 
it suffices to examine the half series $l\ge 0$.
We use \cite[3.4 (6) p.143]{HTF1} with $\mu= -l$, $\nu = \lambda$, i.e.
\beq
\P_\lambda^{-l}(z) = {1\over \Gamma(1+l)}
\left ({1-z\over 1+z}\right )^{l\over 2}
F_l\left({1-z\over 2} \right )\ .
\label{m.38}\endq
where we defined 
\begin{equation}
F_l(z) = F\left(-\lambda,\ 1+\lambda\ ;\ 1+l\ ;\ z \right )
\end{equation}
Thus
\begin{align}
&c_l(z,\ z') = {\gamma_l(\lambda)\over \Gamma(1+l)^2}
\left ({1-z\over 1+z}\right )^{l\over 2}
\left ({1+z'\over 1-z'}\right )^{l\over 2} F_l\left( {1-z\over 2} \right )
F_l\left({1+z'\over 2} \right )\ .
\label{m.39}\end{align}
Simple geometry shows that
\beq
\pm \Re z > 0 \Longleftrightarrow
\left | {1\mp z\over 1\pm z}\right | < 1.
\label{m.40}\endq
Using the discussion in Appendix \ref{appa} (eqs (\ref{ap.40}-\ref{ap.49})),
we make estimates of all the factors occuring in
$c_l(z, z')$ which will be valid even if $l$ is not
an integer, provided $l \ge l_0$ for some $l_0 >0$.

We first let $N$
be the smallest integer $\ge |\Re \lambda|+1$. Then for $l> N$
\begin{align}
&\left |{\Gamma(l-\lambda)\over\Gamma(l+1)}\right | \le
{\Gamma(l+N)\over\Gamma(l+1)} \le (l+N)^{N-1}\le  (2l)^{N-1},\cr
&\left |{\Gamma(l+1+\lambda)\over\Gamma(l+1)}\right | \le
{\Gamma(l+1+N)\over\Gamma(l+1)}\le (2l)^N\ .
\label{m.60}\end{align}
We now set $z = i\tg s $ with $\Im s < 0$ and
$z' = i\tg s' $ with $\Im s' > 0$.
Then
\beq
{1-z\over 1+z} = e^{-2is}, \ \ \ \left |{1-z\over 1+z} \right | < 1, \ \ \
\Re z > 0,\ \ \ {1-z'\over 1+z'} = e^{-2is'}, \ \ \ 
\left |{1+z'\over 1-z'} \right | < 1, \ \ \ \Re z' < 0\ .
\label{m.70}\endq 
To discuss the first hypergeometric function appearing in (\ref{m.39})
we temporarily denote $w = (1-z)/2$ which satisfies 
\beq
\Re w < \haf,\ \ \ 
w = {e^{-is}\over 2\cos(s)},\ \ \
|w| \le {1\over 2|\Im s|}\ .
\label{m.75}\endq
According to (\ref{ap.40}-\ref{ap.49})
\begin{align}
&\left | F_l\left(w \right )
-1\right | \le {1\over l+1} |w|M(w)|1+\Re \lambda|\ch(\pi\Im \lambda),\cr
&\quad\quad M(w) = \sup_{0\le u\le 1}|(1-uw)^{\lambda-1}|\ .
\label{m.80}\end{align}
We have, for $0\le u\le 1$, $\Re uw < \haf$, $|\Arg(1-uw)| < \pi$,
\beq
|(1-uw)^{\lambda-1}| \le |1-uw|^{\Re \lambda -1}e^{\pi|\Im \lambda|}\ .
\label{m.80.1}\endq
with our choices of $\lambda$, $\Re \lambda -1 < -1$. Since $\Re uw < \haf$,
$|1-uw| > \haf$, so that
\beq
M(w) \le 2^{1-\Re \lambda}e^{\pi|\Im \lambda|},
\label{m.81}\endq
\beq
\left | F_l\left({1-z\over 2} \right )
-1\right | \le {1\over (l+1)|\Im s|}
2^{-\Re \lambda} e^{2\pi|\Im \lambda|}\ .
\label{m.82}\endq
An analogous bound, with $s'$ instead of $s$, holds for the second
hypergeometric function occuring in (\ref{m.39}).
Gathering all this shows that
there are positive constants $E$ and $Q$ depending only on $\lambda$
such that
\beq
|c_l(z,\ z')| \le E\left (1+{1\over |\Im s|}\right )
\left (1+{1\over |\Im s'|}\right )(l+1)^Q e^{l(\Im s -\Im s')}\ .
\label{m.83}\endq
Recall again that here $\Im s <0$ and $\Im s'>0$, and that the
bound (\ref{m.83}) only requires $l \ge l_0$ for some $l_0 > 0$,
and the genericity of $\lambda$.

\vskip 0.25 cm
Returning to the case of integer $l$, we see that
the two series
\beq
\sum_{l\in \bZ,\ l \ge 0} c_l(z,z') e^{il(\theta-\theta')}\ \ \
{\rm and}\ \ \ 
\sum_{l\in \bZ,\ l > 0} c_{-l}(z,z') e^{il(\theta'-\theta)}
= \sum_{l\in \bZ,\ l > 0} c_l(z,z') e^{il(\theta'-\theta)}
\label{m.90}\endq
converge absolutely and uniformly on any compact subset of the tubes
\begin{align}
&\{(s,\ s',\ \theta,\ \theta')\ : \Im s <0,\ \ \Im s' >0,\ \
\Im(s'-s-\theta'+\theta) >0\}
\label{m.91}\\
&{\rm and}\cr
&\{s,\ s',\ \theta,\ \theta')\ : \Im s <0,\ \ \Im s' >0,\ \
\Im(s'-s+\theta'-\theta) >0\}
\label{m.92}\end{align}
respectively, and that the limits are holomorphic functions
having boundary values in the sense of tempered distributions
at the real values of $(s,\ s',\ \theta,\ \theta')$.

Hence
$
\sum_{l\in \bZ}c_l(z, z',\lambda) e^{il(\theta-\theta')}
$
 converges to a function holomorphic in the tube
\beq
\rT_{-,+} = \{(s,\ s',\ \theta,\ \theta')\ : \Im s <0,\ \ \Im s' >0,\ \
\Im(s'-s)-|\Im(\theta'-\theta)| >0\}
\label{m.94}\endq
which has a tempered boundary value at the real values of
$(s,\ s',\ \theta,\ \theta')$.
Denoting $u=s+\theta$, $v=s-\theta$, $u'=s'+\theta'$, $v=s'-\theta'$, 
the tube (\ref{m.94}) contains the tube
\beq
\TT_{-,+} =
\{(u,\ v,\ u',\ v')\ : \Im u <0,\ \ \Im v < 0,\ \ \Im u' >0,\ \ \Im v' > 0\}.
\label{m.95}\endq

\subsection{Case of half-odd-integers}
We still consider the series (\ref{cv.10}) 
now assuming that $l\in \half +\bZ$.
\subsection{Positive $l$}
\label{poshalf}
Here $l = n+\half$, with integer $n\ge 0$.
Eqs. (\ref{m.38}) and (\ref{m.39}) remain valid. 
With $\Im s<0$ and $\Im s'>0$, the estimate
(\ref{m.83}) still holds and therefore the series
\beq
\sum_{l \in \half+\bZ,\ l>0}c_l(z, z')e^{il(\theta-\theta')}
\label{g.20}\endq
converges uniformly on every compact of the tube (\ref{m.91})
(hence also of $\rT_{-,+}$ or $\TT_{-,+}$ (see (\ref{m.94}, \ref{m.95})))
to a holomorphic function that has a boundary value in the sense
of tempered distributions at real values of $s$, $s'$, $\theta$,
$\theta'$ (or $u$, $v$, $u'$, $v'$).

\subsection{Negative $l$}
\label{neghalf}
Taking again $l= \half+n$ with integer $n\ge 0$ we consider
\beq
c_{-l}(z,\ z') = \gamma(-l)\P_\lambda^l(z)\P_\lambda^l(-z')
\label{g.30}\endq
and use the formula (obtainable from \cite[3.3.2 (17) p. 141]{HTF1})
\begin{align}
&\P_\lambda^l(z)= {\Gamma(l+\lambda+1)\Gamma(l-\lambda)\over
\pi\Gamma(1+l)} \Bigg [-\sin(\lambda\pi)\left ({1-z\over 1+z}\right )^{l\over 2}
F_l\left ({1-z\over 2}\right )\cr
&+\sin(l\pi) \left ({1+z\over 1-z}\right )^{l\over 2}
F_l\left ({1+z\over 2}\right )
\Bigg ]\ .
\label{k.10}\end{align}
and the identity
\begin{align}
\haf\Gamma(-l-\lambda)\Gamma(\lambda-l+1)
{\Gamma(\lambda+l+1)^2\Gamma(l-\lambda)^2\over \pi^2\Gamma(l+1)^2}
&= {-\gamma_l\over \cos^2(\pi\lambda)\Gamma(l+1)^2}\ .
\label{g.35}\end{align}
We can rewrite
\beq 
c_{-l}(z, z') = \sum_{\veps,\veps' = \pm}
c_{-l,\veps,\veps'}(z,z')
\label{g.36}\endq
where $\veps = \mp$ (resp. $\veps' = \pm$) denotes the choice of
the first or second
term in the bracket of (\ref{k.10}). Thus
\begin{align}
c_{-l,-,-}(z,z')  &= &
{\Gamma(l-\lambda)\Gamma(l+\lambda+1)\sin(\pi l) \sin(\pi\lambda)\over
2\Gamma(l+1)^2 \cos(\pi\lambda)^2}
\left ({1-z\over 1+z}\right )^{l\over 2}
\left ({1-z'\over 1+z'}\right )^{l\over 2}\times\cr 
 && F_l\left ({1-z \over 2}\right )
F_l\left ({1-z' \over 2}\right )\ ,
\label{g.40.1}\\
c_{-l,+,+}(z, z') &= &
{\Gamma(l-\lambda)\Gamma(l+\lambda+1)\sin(\pi l) \sin(\pi\lambda)\over
2\Gamma(l+1)^2 \cos(\pi\lambda)^2}
\left ({1+z\over 1-z}\right )^{l\over 2}
\left ({1+z'\over 1-z'}\right )^{l\over 2}\times\cr
&& F_l\left ({1+z \over 2}\right )
F_l\left ({1+z' \over 2}\right )\ ,
\label{g.40.2}\\
c_{-l,-,+}(z,\z') &=&
{-\Gamma(l-\lambda)\Gamma(l+\lambda+1)\sin^2(\pi\lambda)\over
2\Gamma(l+1)^2 \cos(\pi\lambda)^2}
\left ({1-z\over 1+z}\right )^{l\over 2}
\left ({1+z'\over 1-z'}\right )^{l\over 2}\times\cr
&& F_l\left ({1-z \over 2}\right )
F_l\left ({1+z' \over 2}\right )\ ,
\label{g.40.3}\\
c_{-l,+,-}(z,z') &=&
{-\Gamma(l-\lambda)\Gamma(l+\lambda+1)\sin^2(\pi l)\over
2\Gamma(l+1)^2 \cos(\pi\lambda)^2}
\left ({1+z\over 1-z}\right )^{l\over 2}
\left ({1-z'\over 1+z'}\right )^{l\over 2}\times\cr
&& F_l\left ({1+z \over 2}\right )
F_l\left ({1-z' \over 2}\right )\ .
\label{g.40.4}\end{align}
For a given choice of $\veps$ and $\veps'$ the estimates 
(\ref{m.60}-\ref{m.83}) are readily adapted so that
the series
\beq
\sum_{l= \half+n,\ n\ge 0}
c_{-l,\veps,\veps'}(z,\ z',\ \lambda)e^{-il(\theta -\theta')}
\label{g.41}\endq
converges absolutely to a holomorphic function of 
$(s,\ s',\ \theta,\ \theta')$ in the tube
\beq
\rT_{\veps, \veps'} = \{(s,\ s',\ \theta,\ \theta')\ :\ 
\veps\Im s > 0,\ \ \veps' \Im s' >0,\ \
\Im(\veps' s'+ \veps s)-|\Im(\theta'-\theta)| >0\}
\label{g.42}\endq
as well as in
\beq
\TT_{\veps, \veps'} = 
\{(u,\ v,\ u',\ v')\ :\ \veps\Im u >0,\ \ \veps\Im v > 0,\ \
\veps'\Im u' >0,\ \ \veps'\Im v' > 0\}.
\label{g.43}\endq
This function
has a boundary value at real values of these variables in the sense of
tempered distributions. 

\subsection{Conclusion}
The series (\ref{cv.10}) converges to a distribution $F_0$
which is a finite sum of
boundary values of functions holomorphic in several non-intersecting open
tuboids. Thus $F_0$ is not the boundary value of a function holomorphic
in a single open tuboid. It is possible to verify that this is also
true for the invariant functions of the type given by (\ref{inve}) and
(\ref{invo}). It might be asked if $F_0$ (or one of its siblings)
could not still have a tuboid of analyticity beyond what follows
from the above proofs of convergence.
However in the following Sect.~\ref{incomp} a general lemma will show that
analyticity is incompatible with the simultaneous requirements
of locality (C1), invariance (C3) and Klein-Gordon equation (C6).
The proof of convergence given in this subsection also works for
more general (non-invariant) series of the form (\ref{wig}) provided
the $A_l,\ ... C_l^*$ are polynomially bounded in $l$.

\newpage

\section{Incompatibility of analyticity with some other requirements}
\label{incomp}
In this section the following lemma will be proved:
\begin{lemma}
\label{incomp1}
A two-point function $F$ on $\wt{dS_2}\times \wt{dS_2}$ that simultaneously
satisfies (C3) Invariance under $\slr$, 
(C6) Klein-Gordon equation, and (C9) Local analyticity (see Sect. \ref{proof}),
vanishes on $\RR$ (i.e. $F(x,x')=0$ whenever $x$ and  $x'$ are
space-like separated).
\end{lemma}
An example of a non-zero two-point function satisfying these requirements
is the canonical commutator (\ref{coeff2}).
As an obvious corollary of this lemma,
\begin{lemma}
\label{incomp2}
A two-point function $F$ on $\wt{dS_2}\times \wt{dS_2}$ that simultaneously
satisfies (C1) Locality, (C3) Invariance under $\slr$, 
(C6) Klein-Gordon equation, and (C8) Analyticity (see Sect. \ref{proof}),
is equal to 0.
\end{lemma}

{\bf Proof}. Suppose $F$ satisfies the conditions (C3), (C6), and (C9).
As in Sect. \ref{proof},
let $\RR$ denote the (open, connected) set of space-like separated points in
$\wt{dS_2}\times \wt{dS_2}$. and 
$\NN$ a complex open connected neighborhood of $\RR$ in which $F(\wtx, \wtx')$
and $F(\wtx', \wtx)$ have
a common analytic continuation. We denote $F_+$
this analytic continuation.
For any pair $k = (\wt w,\ \wt w')$ we denote, by abuse of notation,
$\wt w\cdot \wt w'$ or also $\psi(k)$
the scalar product of the projections $w$ and $w'$
of $\wt w$ and $\wt w'$ into the complex
Minkowski space $M_3^{(c)}$ (i.e. with an abuse of notation,
$\psi(k) = -1 -\half (w-w')^2$). In particular $\NN$ contains the subset
\beq
E_{\veps,\ \eta} = \{(t,\ \theta+iy),\ (t',\ \theta'))\ :\ 
t = t' = \theta' = 0,\ \ \veps < \theta < 4\pi-\veps,\ \ \ 
|y| < \eta \}\ ,
\label{ap.5}\endq
where $\veps > 0$ and $\eta >0$ must be chosen small enough.
Note that for points of the form (\ref{ap.5}),
\beq
\z = -\cos(\theta+iy) = -\cos(\theta)\ch(y) + i\sin(\theta)\sh(y)\ .
\label{ap.6}\endq

Let $k_0 = (\wt w_0,\ \wt w'_0) \in \NN$ be such that
$\z_0 = \psi(k_0) \not=  \pm 1$. 
There exist  open neighborhoods 
$U_1\subset \subset U_2 \subset \subset\NN$ of $k_0$, and an open
neighborhood $W_0$ of the identity in the group $\slc$
such that, for all $g\in W_0$ and $k \in U_1$, $gk \in U_2$ and
$F_+(gk) = F_+(k)$ (since this holds for real $g$). Moreover we suppose
$U_2$ small enough that the restriction to $U_2$ of the projection
$\pr\times\pr$ is an isomorphism, and also that
$k \mapsto \sqrt{1-\psi(k)^2}$ can be defined as a holomorphic function
on $U_2$ (in particular $\psi(k) \not= \pm 1$ for all $k\in U_2$).

We will prove\footnote{These arguments are special cases of more general
well-known facts. See e.g. \cite{Hall-Wightman}, \cite {Hepp}.}
that there is an
open neighborhood $V_0 \subset \subset U_1$ of $k_0$, and
a function $f_0$ holomorphic on $\psi(V_0)$ such that
$f_0(\psi(k)) = F_+(k)$ for all $k\in V_0$.
To do this we adopt the simplifying notation whereby if $\wt{t} \in
\wt{dS_2}^{(c)}$ then $t$ denotes $\pr \wt{t}$ and conversely
if $t\in dS_2^{(c)}$ then $\wt{t}$ denotes $\pr^{-1} t$.
For any $k= (\wt{w},\ \wt{w'}) \in U_2$ we construct
a complex Lorentz frame $(e_0(k),\ e_1(k), e_2(k))$ as follows:
\begin{align}
&e_0(k) = iw\ ,\cr
&e_1(k) = \alpha w + \beta w',\ \ e_0(k)\cdot e_1(k) =0,\ \ \ 
e_1(k)^2 = -1\ ,\cr
&e_2(k) = e_0(k) \times e_1(k)\ ,\ \ {\rm i.e.}\ \
e_2(k)^\mu = -\veps^{\mu\nu\rho} e_0(k)_\nu\, e_1(k)_\rho\ .
\label{ap.10}\end{align}
Denoting $\z = w\cdot w' = \psi(k)$, this implies
\beq
\alpha = \beta\z,\ \ \ \beta = {\pm 1\over \sqrt{1-\z^2}}\ ,\ \ 
\hbox{and we choose}\ \ \beta = {1\over \sqrt{1-\z^2}}\ ,
\label{ap.15}\endq
where $\sqrt{1-\z^2}$ denotes the determination of $\sqrt{1-\psi(k)^2}$
mentioned above. This implies
\beq
w = -ie_0(k),\ \ w' = \sqrt{1-\z^2}\,e_1(k) + i\z\,e_0(k)\ .
\label{ap.20}\endq
For any $\zeta$ sufficiently close to $\z_0$ let
$h(\z)= (\wt{v(\z)},\ \wt{v'(\z)})$
be defined by
\beq
v(\z) = -ie_0(k_0),\ \ v'(\z) = \sqrt{1-\z^2}\,e_1(k_0) + i\z\,e_0(k_0)\ .
\label{ap.25}\endq
It is clear that
$v(\z)^2 = v'(\z)^2 = -1$, $v(\z)\cdot v'(\z) = \z$, and
$e_j(h(\z)) = e_j(k_0)$ for $j= 0,\ 1,\ 2)$.
If $k = (\wt{w},\ \wt{w'})$ is close to $k_0$ and $\psi(k) = \z$, there exists
an element $g(k)$ of $\slc$  close to the identity, such that
$k = g(k)h(\z)$. This element projects onto the unique Lorentz transformation
$\Lambda(k)$ such that $e_j(k) = \Lambda(k)e_j(h(\z)) = \Lambda(k)e_j(k_0)$
for $j= 0,\ 1,\ 2)$.
We have therefore $F_+(k) = F_+(h(\z))$, i.e. a holomorphic function of $\z$.
We have now shown that every $k_0 \in \NN$ such that $\psi(k_0)\not= \pm 1$
has an open neighborhood $V_0$ such that $F_+(k) = f_0(\psi(k))$ for
all $k\in V_0$, where $f_0$ is holomorphic in $\psi(V_0)$.

If $k_1$ is another point of $\NN$ such that $\psi(k_1)\not= \pm 1$, 
and $V_1$, $f_1$  are the analogous objects, and if $V_0$ and
$V_1$ overlap, it is clear that $f_1$ is an analytic continuation of $f_0$.
Thus for any compact arc contained in $\NN$ from $k_0$ to $k_2$, $f_0$ can be
analytically continued along the image under $\psi$ of that arc, provided
this image avoids the points $\pm 1$.

\begin{figure}[ht]
\begin{center}
\includegraphics{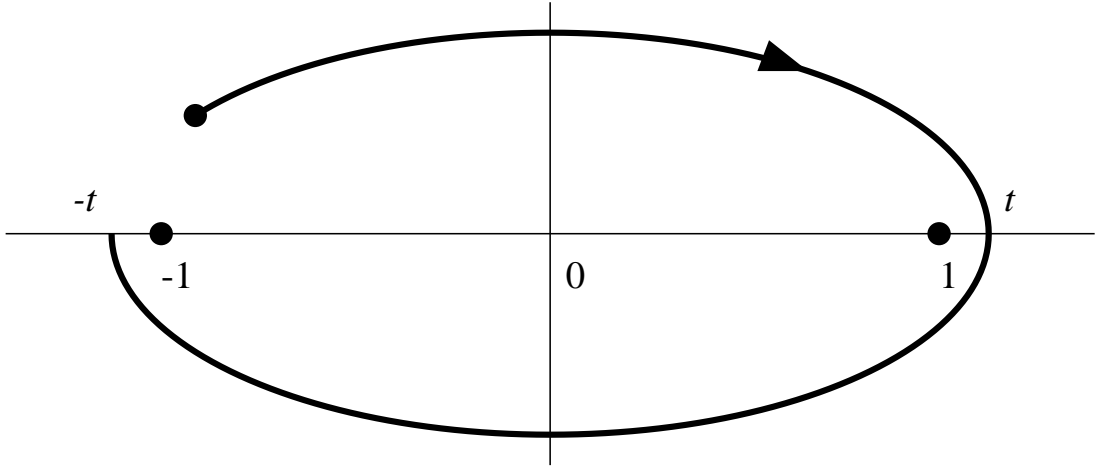}
\caption{The arc of ellipse described by $\z$}
\label{ellfig}
\end{center}
\end{figure}

Since $F$ satisfies the Klein-Gordon equation in $x$ and in $x'$, it follows,
by a well-known calculation, that the $f=f_0$ obtained by the above procedure
at a point $k_0$ (with $\psi(k_0) \not= \pm 1$) must be a solution of
the Legendre equation (\ref{s.19}) with $l=0$. 
By the general theory of such equations, $f$ may be analytically
continued along any arc in the complex plane which avoids the points
$\pm 1$. As before, two linearly independent solutions of the equation are
\beq
\P_\lambda(\z) = \P_\lambda^0(\z)  = F \left (-\lambda,\ \lambda+1\ ;\ 1\ ;\ {1-\z\over 2}
\right )\ . 
\label{p.116}\endq
 and
$\cP_\lambda(\z) = \cP_\lambda^0(\z) = \P_\lambda^0(-\z)$. 
Recall that $\P_\lambda$ is holomorphic in the cut-plane with a cut along
$(-\infty,\ -1]$ and is singular at $-1$ : according to \cite[p. 164]{HTF1}
it has a logarithmic singularity at $-1$. Hence 
$\cP_\lambda$ is holomorphic in the cut-plane with a cut along
$[1,\ \infty)$ and has a logarithmic singularity at $1$.
It follows from \cite[(36) pp. 132-133]{HTF1} that $Q_\nu^\mu$ is
holomorphic in $\bC\setminus (-\infty,\ 1]$, in particular on $(1,\ \infty)$.
By \cite[(10) p. 140]{HTF1}, for real $t > 1$,
\beq
\cP_\lambda(t+i0) - \cP_\lambda(t-i0) =
-2i\sin(\lambda\pi) \P_\lambda(t)\ ,\ \ \ t > 1\ .
\label{p.117}\endq
We must have
$
f(\z) = a \P_\lambda(\z) + b\cP_\lambda(\z) \ ,
$
where $a$ and $b$ are constants. We specialize $k_0$ as
\beq
k_0 = ((t_0 = 0,\ \theta_0 = 2\veps),\ (t'_0 = 0,\ \theta'_0 = 0))\ ,
\label{p.130}\endq
where $0< \veps $ is as in (\ref{ap.5}), and we suppose $\veps < \pi/8$.
We denote
\beq
G(\theta +iy) = F_+((t=0,\ \theta +iy),\ (t'=0,\ \theta'=0)).
\label{p.131}\endq
As the restriction of $F_+$ to the set $E_{\veps,\ \eta}$, $G$ is holomorphic 
in the rectangle $\veps < \theta < 4\pi-\veps$, $|y| <\eta$.
We consider an arc (actually a straight line) $\gamma_y$ lying in
$E_{\veps,\ \eta}$, given by
\beq
\theta \mapsto \gamma_y(\theta) = ((t=0,\ \theta +iy),\ (t'=0,\ \theta'=0)).
\label{p.140}\endq
Here $y$ is real with $|y| \le \tau$ and $\theta$ varies in the interval
$[2\veps,\ 2\pi]$. We also require $0 < \tau < \eta$ to be small enough
that the starting point $\gamma_y(2\veps)$ be always contained in
the neighborhood $V_0$ of $k_0$ where the function $f = f_0$ is
initially defined. Let $y$ be fixed with $0 < y <\tau$. As
$\theta$ varies in $[2\veps,\ 2\pi]$,
$\z = \psi(\gamma_y(\theta)) = -\cos(\theta+iy)$ runs along an arc
of an ellipse with foci at $\pm 1$, starting in the upper half-plane,
crossing the real axis at $t = \ch(y)$, and returning through the
lower half-plane to $-t-i0$ (see Fig.~\ref{ellfig}).
Along this arc, $f$ can be analytically
continued; starting as $f(\z) = a\P_\lambda(\z) + b \cP_\lambda(\z)$
it becomes, after crossing the real axis at $t$, equal to
$[a -2ib\sin(\lambda\pi)] \P_\lambda(\z) + b\cP_\lambda(\z)$ (as a consequence
of (\ref{p.117})). Thus at the end point,
\beq
f(-t-i0) = G(2\pi+iy) = [a -2ib\sin(\lambda\pi)] \P_\lambda(-t-i0)
+ b \cP_\lambda(-t-i0)\ .
\label{p.145}\endq
Since $|G(2\pi+iy)|$ is bounded uniformly in $y$, while
$|\P_\lambda(-t-i0)| \rightarrow \infty$ as $t \rightarrow 1$, we must
have 
\beq
a - 2ib\sin(\lambda\pi) = 0.
\label{p.146}\endq
Repeating the argument with $y<0$ (the arc of ellipse starts in the
lower half-plane and finishes in the upper half-plane) we now get
\beq
a + 2ib\sin(\lambda\pi) = 0.
\label{p.147}\endq
This implies $f=0$, and therefore that $F_+$ vanishes in an open
complex neighborhood of $k_0$. Hence $F_+ = 0$, hence $F$
vanishes on $\RR$.

There are examples of 2-point functions that satisfy any three out
of the four conditions (C1), (C3), (C6), and (C8).
For instance let
\begin{align}
F_2(x,\ x', \lambda) = c_0(z,z', \lambda)
+ \sum_{l \in \LL,\ l> 0}
c_l(z,z',\lambda)  [ e^{il(\theta-\theta')}+ e^{-il(\theta-\theta')}
\big ]\ ,
\label{n.30}\end{align}
where $z=i\sh t$, $z'=i\sh t'$, and
\beq
c_l(z,z', \lambda) =
\gamma_l(\lambda)\P_\lambda^{-l}(z)\P_\lambda^{-l}(-z')\ .
\label{n.31}\endq
Then $F_2$ satisfies (C1), (C6) as well as canonicity (up to a constant factor)
and positivity.
It also satisfies (C8) and is the boundary value of a function holomorphic
in the tuboid $\rT_{-,+}$ (see (\ref{m.94})) (or $\TT_{-,+}$ (see (\ref{m.95}))).
{\em But $F_2$ is not invariant under $\slr$.}

\section{Concluding remarks : absence of  the Gibbons-Hawking temperature}
We have seen that an invariant local two-point function satisfying
the Klein-Gordon equation cannot have
the property of analyticity. As a consequence, if such a function is used,
a geodesic observer on the double covering of the two-dimensional de Sitter
space-time will never detect a thermal bath of particles, even though he or she
cannot distinguish the global topological structure of the space time manifold
in any other way! Let us summarize some well-known facts  in the case of
the two-dimensional de Sitter spacetime. The wedge
\beq
U = \{ x\in dS_2, \ \ |x^1| \leq 1, \ \ x^2 \geq 0 \}.
\endq
is invariant under the one-parameter subgroup of $\slr$ given by
\beq
t \mapsto g(t) = \left(
\begin{array}{ccc}
 \cosh t & 0 & \sinh t \\
 0 & 1 & 0 \\
 \sinh t & 0 & \cosh t
 \\
\end{array}
\right).
\label{autom}\endq
As $t$ varies in $\bR$ the point
\beq
x(t,r)= \left(
\begin{array}{c}
 \sqrt{1-r^2} \, \sinh t\\
 r \\
 \sqrt{1-r^2}\,  \cosh t \\
\end{array}
\right) = g(t)x(0,r),\ \ \ {\rm where}\ \ -1<r<1,
\label{orbits}\endq
describes an orbit of this subgroup. This is a branch of a hyperbola,
and, in the case $r=0$, it is a geodesic of $dS_2$, which can be
regarded as the world-line of a ``geodesic observer''. The orbits
(\ref{orbits}) are restrictions of complex curves in $dS_2^{(c)}$
obtained by letting $t$ vary in $\bC$. If $t = a+ib$ with $a\in \bR$
and $b = \pi\ \mod 2\pi$, $x(t,\ r)$ describes another real branch of
the complex hyperbola.

Suppose now that $W(z_1,z_2)$ is a maximally analytic two-point function in
the complex de Sitter hyperboloid; let the first point $z_1 = x(t_1,r_1)$
be fixed in  the wedge $U$  and let 
the second point vary on the complex hyperbola
$z_2 = x(t_2,r_2)$ ($t_2$ is a complex time variable).
It easy to check \cite{bm} that the two-point function,
\begin{equation}
W( z_1, z_2)  = W ( -\sqrt{1-r_1^2}\sqrt{1-r_2^2} \cosh(t_1-t_2) - r _1r_2)
\end{equation}
as a function of the complex variable $t_2$ is holomorphic on the strip
$\{ 0 < \Im (t_2) <2 \pi \}$ and that the boundary values satisfy the
KMS condition 
\begin{eqnarray}
\lim_{\epsilon\to 0}{W}(x_1(t_1,r_1), x_2 (t_2+i \epsilon, r_2))=
{\cal W} (x_1,x_2), \\
\lim_{\epsilon\to  0}{W}(x_1(t_1,r_1), x_2 (t_2+2 \pi i - i \epsilon, r_2))=
{\cal W} (x_2,x_1), 
\end{eqnarray}
These relations are usually interpreted by saying that the geodesic observer
perceives a thermal bath of particles at temperature $T= 1/2\pi$.
Analyticity and locality actually suffice for these relations to hold,
without supposing invariance (\cite{bem}).
But these properties disappear in the case of the two-sheet covering
$d\wt{S}_2$ of the two-dimensional de Sitter space.

To understand this in more detail, let us again consider
$F_2(x,\ x', \lambda)$ given by (\ref{n.30}) and (\ref{n.31}),
and
\beq
F_0(x,\ x', \lambda) = \sum_\LL c_l(z,\ z',\ \lambda)e^{il(\theta-\theta')}\ ,
\label{b.15}\endq
\beq
F_1(x,\ x', \lambda) = \haf F_0(x,\ x', \lambda) +
\haf F_0(\tau_1 x',\ \tau_1 x, \lambda) = 
\sum_{l\in \LL} \haf c_l(z,\ z',\ \lambda)
\Big [ e^{il(\theta-\theta')} + e^{-il(\theta-\theta')} \Big ]\ .
\label{e.20}\endq
Here $\LL = \haf\bZ$. We use the same notations as in Sect. \ref{conver}.
Recall that $F_0$ and $F_1$ are invariant but not
analytic,
and $F_2$ is analytic but not invariant. $F_1$ and $F_2$ are local,
$F_0$ is not. Let $F_{j,\rm even}$ (resp. $F_{j,\rm odd}$) 
denote (for $j = 0,\ 1,\ 2$) the sum of the correspoding series over
integer (resp. non-integer) values of $l$. Since for integer $l$
$c_l(z,\ z',\ \lambda) = c_{-l}(z,\ z',\ \lambda)$ (see Subsect. \ref{convint}),
we have $F_{2,\rm even} = F_{0,\rm even}$. This is an invariant and analytic
2-point function, local in the sense of $dS_2$, and in fact
(see Sect. \ref{seccomminv}) it coincides with the Bunch-Davies function. Hence
it possesses the analyticity along complex hyperbolae discussed at the
beginning of this section.

To understand the behavior of the $F_{j,\rm odd}$, we study
their limits as $\lambda$ tends to 0.
If $\lambda \rightarrow 0$, $c_l(z,\ z',\ \lambda)$
tends to a well-defined limit provided $l \not= 0$.
If $l \in \half+\bZ$ and $l>0$,
\beq
c_l(z,\ z',\ 0)  \bydef \lim_{\lambda\rightarrow 0} c_l(z,\ z',\ \lambda)
= {1\over 2l}\left ({1-z\over 1+z}\right )^{l\over 2}
\left ({1+z'\over 1-z'}\right )^{l\over 2}
= {1\over 2l} e^{il(s'-s +i\veps)}\ ,
\label{h.10}\endq
\beq
c_l(z,\ z',\ 0)e^{il(\theta-\theta')}= {1\over 2l} e^{il(v'-v)}\ ,\ \ 
c_l(z,\ z',\ 0)e^{-il(\theta-\theta')}= {1\over 2l} e^{il(u'-u)}\ .
\label{h.16}\endq
For the same $l \ge \half$, we find
\beq
c_{-l}(z,\ z',\ 0)e^{-il(\theta-\theta')} =
-{1\over 2l} e^{il(v-v')}\ ,\ \ c_{-l}(z,\ z',\ 0)e^{il(\theta-\theta')} =
-{1\over 2l} e^{il(u-u')}\ .
\label{h.28}\endq
\beq
F_{0,{\rm odd}}(x,\ x',\ 0) = 
\haf \log \left ( {1+ e^{i(v'-v+i\veps)/2}
\over 1- e^{i(v'-v+i\veps)/2}} \right )
-\haf \log \left ( {1+ e^{i(v-v'+i\veps)/2}
\over 1- e^{i(v-v'+i\veps)/2}} \right )\ .
\label{h.45}\endq
The first (resp. second) term is the boundary value term of a function
holomorphic in the tuboid $\{\Im (v'-v) >0\}$ (resp $\{\Im (v'-v) <0\}$).
$F_1$ is the sum of the boundary values of four functions
analytic in two pairs of opposite tuboids:
\begin{align}
F_{1,{\rm odd}}(x,\ x',\ 0) &=  
{1\over 4} \log \left ( {1+ e^{i(u'-u+i\veps)/2}
\over 1- e^{i(u'-u+i\veps)/2}} \right )
-{1\over 4} \log \left ( {1+ e^{i(u-u'+i\veps)/2}
\over 1- e^{i(u-u'+i\veps)/2}} \right )\cr 
&+{1\over 4} \log \left ( {1+ e^{i(v'-v+i\veps)/2}
\over 1- e^{i(v'-v+i\veps)/2}} \right )
-{1\over 4} \log \left ( {1+ e^{i(v-v'+i\veps)/2}
  \over 1- e^{i(v-v'+i\veps)/2}} \right )\ .
\label{h.46}\end{align}
If $x' = (t',\ \theta')$ and $\theta$ are  fixed real, then
$t\mapsto F_{0,{\rm odd}}((t,\ \theta),\ x')$ is not the boundary value of
a function holomorphic in a strip of the upper half-plane. The same is true
for $F_{1,{\rm odd}}((t,\ \theta),\ x')$.
On the other hand
\begin{align}
F_{2,{\rm odd}}(x,\ x',\ 0) 
&=\haf \log \left ( {1+ e^{i(v'-v+i\veps)/2}
\over 1- e^{i(v'-v+i\veps)/2}} \right ) +
\haf \log \left ( {1+ e^{i(u'-u+i\veps)/2}
\over 1- e^{i(u'-u+i\veps)/2}} \right )\cr
&= -\haf \log \left (- {\sin\left({v'-v+i\veps\over 4}\right )
\sin\left({u'-u+i\veps\over 4}\right )\over
\cos\left({v'-v+i\veps\over 4}\right )
\cos\left({u'-u+i\veps\over 4}\right )}\right )
\label{h.75}\end{align}
is the boundary value of a function of $u,\ v,\ u',\ v',$
holomorphic in the tube
\beq
\{(u,\ v,\ u',\ v')\ :\ \Im(u'-u) > 0,\ \Im(v'-v) > 0\}\ ,
\label{h.80}\endq
which, of course, can be continued in a larger domain. In fact starting
from any point in the tube (\ref{h.80}), the function can be analytically
continued along any arc which does not contain any point
such that $u-u'\in 2\pi\bZ$ or $v-v' \in 2\pi\bZ$. In particular
$F_{2,{\rm odd}}(x,\ x',\ 0)$ is analytic at all real points such that
$u-u' \notin 2\pi\bZ$ and $v-v' \notin 2\pi\bZ$.
However we have from (\ref{h.75})
\beq
G(x,\ x') \bydef \exp(-2F_{2,{\rm odd}}(x,\ x',\ 0)) = 
{\cos\left({s'-s\over 2}\right )-\cos\left({\theta'-\theta\over 2}\right )
\over
\cos\left({s'-s\over 2}\right )+\cos\left({\theta'-\theta\over 2}\right )}\ .
\label{h.90}\endq
Fixing $\theta$ and $\theta'$ real and $s'=0$, and substituting
\beq
\cos(s/2) = [\half (1+\cos(s))]^\haf =
\left [\half \left (1+{1\over\ch(t)}\right )\right ]^\haf\ ,
\label{h.100}\endq
(\ref{h.90})  becomes
\beq
G(x,\ x') = {\sqrt{\half (1+\ch(t))}
- \sqrt{\ch(t)}\cos\left({\theta'-\theta\over 2}\right ) \over
\sqrt{\half (1+\ch(t))}
+ \sqrt{\ch(t)}\cos\left({\theta'-\theta\over 2}\right )}\ .
\label{h.105}\endq
If $\cos((\theta'-\theta)/2) \not= 0$,
this is singular in $t$ when $\ch(t)$ vanishes, i.e.
$it \rightarrow \pi/2\ \mod \pi$, so that $F_2$ is not analytic in $t$
in the strip $0< \Im t < \pi$. This does not prove, but makes it very
likely that the same holds for $F_2(x,\ x',\ \lambda)$ with
$\lambda \not= 0$.

\vskip 20 pt
\noindent{\bf \large Acknowlegdements.} U.M. thanks the IHES where the main body of this work has been done.

\appendix
\section{Appendix. Estimations for hypergeometric functions}
\label{appa}
We reproduce here for completeness some estimates from
\cite[2.3.2 pp 76-77]{HTF1}.
Recall (see e.g. \cite[5.6(ii)]{NIST})
\begin{align}
|\Gamma(x+iy)| \le |\Gamma(x)|,
\label{ap.0} \ \ \ 
|\Gamma(x+iy)| &\ge {1\over \sqrt{\ch y}}|\Gamma(x)|,\ \ \ x\ge \haf\ .\end{align}
Define
\begin{align}
&\rho_{n+1} (a,\ b,\ ;\ c\ ;\ z) = F(a,\ b,\ ;\ c\ ;\ z) -1 - {ab\over c}z - ...
- {(a)_n(b)_n\over (c)_n n!}z^n\cr
&={\Gamma(c)\Gamma(a+n)z^{n+1}\over \Gamma(b)\Gamma(c-b)\Gamma(a)n!}
\int_0^1 ds \int_0^1 dt\,t^{b+n}(1-t)^{c-b-1}(1-s)^n(1-stz)^{-a-n-1}\ .
\label{ap.40}\end{align}
Let $a=\alpha+i\alpha'$, $b=\beta+i\beta'$, $c = \gamma +i\gamma'$. Then
\beq
|\rho_{n+1}| \le \mu(n)\,|z|^{n+1} |c|^{-\beta} \gamma^{-\beta-n-1}\ ,
\label{ap.41}\endq
where it is assumed that $|\arg(1-z)|<\pi$, $\gamma >\beta$,
$n+\beta> 0$, $|\arg c|<\pi -\veps$, $\gamma>0$ sufficiently large,
$n$ sufficiently large. $\mu(n)$ depends on $n$, $a$, $b$, $z$.

Example: $n=0$
\begin{align}
&\rho_1(a,\ b,\ ;\ c\ ;\ z) =
{\Gamma(c)z\over \Gamma(b)\Gamma(c-b)}\,I\ ,\cr
&I = \int_0^1 ds \int_0^1 dt\,t^{b}(1-t)^{c-b-1}(1-stz)^{-a-1}\ .
\label{ap.45}\end{align}
We assume $\beta+1 >0$, $\gamma-\beta >0$, $\gamma > 0$,
and that $|1-z| > \veps$, $|\arg(1-z)|<\pi-\veps$ for some $\veps>0$.
In this case the modulus $|I|$ of the integral is bounded by
\beq
|I| \le M(z)\int_0^1 t^\beta (1-t)^{\gamma-\beta -1}\,dt\
\label{ap.46}\endq
with
\beq
M(z) = \sup_{0\le u\le 1}|(1-uz)^{-a-1}|.
\label{ap.47}\endq
\beq
|I| \le M(z) {\Gamma(\beta+1)\Gamma(\gamma-\beta)\over
\Gamma(\gamma+1)}.
\label{ap.48}\endq
Hence using (\ref{ap.0})
\beq
|\rho_1(a,\ b,\ ;\ c\ ;\ z)| \le {|z|M(z)|\beta|
\sqrt{\ch (\pi\beta')\ch\pi(\gamma'-\beta')}\over
\gamma}\ .
\label{ap.49}\endq


\begin{thebibliography}{30}    
\bibitem{hawking1} S. W. Hawking,  
Nature. 248 (5443): 30-31  (1974).
\bibitem{hawking2} S. W. Hawking
Commun. Math. Phys., 43 ( pp. 199-220, 1975).
\bibitem{CM} G. Chibisov and V. Mukhanov 
JETP Lett, 33, No.10, 532 (1981)
\bibitem{Luminet} J-P. Luminet, Scholarpedia, 10(8):31544 (2015).
\bibitem{LR} M. Lachieze-Rey and  J-P. Luminet, Phys. Rept. 254,  135 - 214 (1995).
\bibitem{sch} J. Schwinger, Phys. Rev. D128, 2425 (1962).
 \bibitem{thir}  W. Thirring,   Annals of Physics. 3: 91  (1958).
 \bibitem{kla} B. Klaiber,   Lect. Theor. Phys. 10A: 141 (1968).
\bibitem{Epstein1}
  H.~Epstein and U.~Moschella,
  JHEP {\bf 1605} (2016) 147
\bibitem{difrancesco}
 P. Di Francesco,  P. Mathieu and  D. S\'en\'echal, Conformal Field Theory, Springer, New York, 1996.

  \bibitem{Epstein2}
  H.~Epstein and U.~Moschella,
  In preparation.

 \bibitem{bmds2} H.~Epstein and U.~Moschella,
  Int.\ J.\ Mod.\ Phys.\ A {\bf 33}, no. 34, 1845009 (2018)
  [arXiv:1901.10874 [hep-th]].
  
  \bibitem{Bunch} 
  T.~S.~Bunch and P.~C.~W.~Davies,
  Proc.\ Roy.\ Soc.\ Lond.\ A {\bf 360}, 117 (1978).
  doi:10.1098/rspa.1978.0060

  \bibitem{Chernikov} 
  N.~A.~Chernikov and E.~A.~Tagirov,
  Ann.\ Inst.\ H.\ Poincare Phys.\ Theor.\ A {\bf 9}, 109 (1968).
   \bibitem{hawking} G.~W.~Gibbons and S.~W.~Hawking,
  Phys.\ Rev.\ D {\bf 15}, 2738 (1977).

 
  \bibitem{bm}
  J.~Bros and U.~Moschella,
  Rev.\ Math.\ Phys.\  {\bf 8}, 327 (1996)
 
  
  \bibitem{bgm} 
  J.~Bros, U.~Moschella and J.~P.~Gazeau,
  Phys.\ Rev.\ Lett.\  {\bf 73}, 1746 (1994).

  
   \bibitem{bem} 
     J.~Bros, H.~Epstein and U.~Moschella,
     ``Analyticity Properties and Thermal Effects for General
Quantum Field Theory on de Sitter Space-Time,''
   Commun.\ Math.\ Phys.\  {\bf 196}, 535 (1998)



\bibitem{kayw} B. Kay, S. Wald Phys. Rept. 207, 49-136 (1991). 

\bibitem{HTF1}
A.~Erdelyi, W.~Magnus, F.~Oberhettinger, F.~G.~Tricomi:
{\it Higher Transcendental Functions}, Vol.~1.
New York: McGraw-Hill, 1953
 \bibitem{Birrell}
N. D. Birrell,  P. C. W. Davies, Quantum Fields in Curved Space,
Cambridge University Press, 1984. 

 \bibitem{ms}
  U.~Moschella and R.~Schaeffer,
  JCAP {\bf 0902}, 033 (2009)

  
  \bibitem{ms2} 
  U.~Moschella and R.~Schaeffer,
  AIP Conf.\ Proc.\  {\bf 1132}, 303 (2009)



\bibitem{sw} R. Streater and A.S. Wightman, PCT, Spin and Statistics and all that, Princeton University Press (2000).






 

  
 \bibitem{spindel} 
  C.~Schomblond and P.~Spindel,
  Ann.\ Inst.\ H.\ Poincare Phys.\ Theor.\  {\bf 25}, 67 (1976).
   \bibitem{allen} 
    B.~Allen,
  Phys.\ Rev.\ D {\bf 32}, 3136 (1985).
  \bibitem{mottola}
  E.~Mottola, 
  Phys.\ Rev.\ D {\bf 31}, 754 (1985).
    \bibitem{tach} 
  H.~Epstein and U.~Moschella,
  Commun.\ Math.\ Phys.\  {\bf 336}, no. 1, 381 (2015)
\bibitem{Akhmedov} 
  E.~T.~Akhmedov, K.~V.~Bazarov, D.~V.~Diakonov, U.~Moschella, F.~K.~Popov and C.~Schubert,
  arXiv:1905.09344 [hep-th].


\bibitem{Hall-Wightman}
D.~Hall and A.S.~Wightman:
A theorem on invariant analytic functions with applications to
relativistic quantum field theories.
Mat. Fys. Medd. Dan. Vid. Selsk. {\bf 31} No. 5, 1957

\bibitem{Hepp}
K.~Hepp:
Klassische komplexe Liesche Gruppen und kovariante analytische Funktionen.
Math. Annalen {\bf 152}, 149-158, 1963.
  \bibitem{NIST} NIST Digital Library of Mathematical Functions,https://dlmf.nist.gov/


  
  
  \end{thebibliography}
\end{document}